\documentclass[11pt]{article}
\usepackage{amsmath,amssymb}
\usepackage[nosort]{cite}
\usepackage[margin=1in]{geometry}
\usepackage{graphicx}

\makeatletter \@addtoreset{equation}{section} \makeatother

\makeatletter
\newcommand\blfootnote{\xdef\@thefnmark{}\@footnotetext}
\makeatother

\newcommand{\beq}{\begin{equation}}
\newcommand{\eeq}{\end{equation}}
\newcommand{\bea}{\begin{eqnarray}}
\newcommand{\eea}{\end{eqnarray}}

\input{tcilatex}

\begin{document}


\begin{titlepage}


\vspace*{.2cm}

\begin{center}

\vspace*{1cm}

 {\LARGE\bf  Studies on 1/4 BPS and 1/8 BPS geometries}

\vspace*{.5cm%
}

\vspace*{1.5cm}

 {\bf Hai Lin }

\blfootnote{E-mail: \vtop{ \hbox{%
 hai.lin@usc.es}}}

\vspace*{1cm}
\small{

{\it\large Department of Particle
Physics, Faculty of Physics, \\
University of Santiago de
Compostela,
15782, Santiago de
Compostela, Spain \\} }
\end{center}

\vspace*{.2cm}%

\vspace*{2cm} \large
\begin{abstract}

\vspace*{0.5cm}

{\large We analyze more explicitly four sectors of 1/4 BPS
geometries and 1/8 BPS geometries, corresponding to BPS states in
${\cal N}$=4 SYM, constructed previously. These include the states
with several $SO(6)$ angular momenta as well as those with $SO(4)$
$AdS$ spins. We also discuss their relations to the dual gauge
theory.}

\end{abstract%
}

\end{titlepage}

\setcounter{tocdepth}{2}

\large

\section{Introduction}

In this paper, we study certain sectors of 1/4 BPS and 1/8 BPS geometries in
type IIB string theory.

The 1/2 BPS geometries corresponding to a class of 1/2 BPS states in $%
\mathcal{N}$=4 SYM were studied in \cite{Berenstein:2005aa} -\cite%
{Lin:2004nb}. Many geometric and topological aspects of the geometries were
discussed in 
e.g. \cite{Berenstein:2005aa} -\cite{Balasubramanian:2007zt}. Geometries
corresponding to Wilson loops, surface operators, and defect operators were
constructed and discussed in e.g. \cite{Yamaguchi:2006te} -\cite%
{D'Hoker:2007fq}, \cite{Gukov:2008sn} -\cite{Lin:2005nh}.

On the other hand, 1/4 BPS and 1/8 BPS geometries were studied by e.g. \cite%
{Chong:2004ce} -\cite{Liu:2004ru}. For the 1/4 BPS and 1/8 BPS geometries,
we will in this paper, continue to analyze more detailed explicit
geometries, and their relation to the dual gauge theory. We will also study
separately the four different sectors of geometries.

The organization of this paper is as follows. In section \ref{sec:1-4_JJ},
we analyze geometries with angular momenta $J_{1},J_{2}~$in $S^{5}~$%
directions. In section \ref{sec:1-8_JJJ}, we analyze geometries with $%
J_{1},J_{2},J_{3}$ in $S^{5}~$directions. In section \ref{sec:1-4_SJ}, we
analyze the geometries with spin $S_{1}~$in $AdS_{5}~$and $J~$in $S^{5}~$%
directions respectively. In section \ref{sec:1-8_SSJ}, we analyze geometries
with spins $S_{1},S_{2}~$in $AdS_{5}~$and $J~$in $S^{5}~$directions
respectively. Finally, we briefly conclude in section \ref{sec:discussion}.

\section{1/4 BPS geometries with $J_{1},J_{2}$}

\label{sec:1-4_JJ} 

\subsection{General ansatz}

\label{sec:1-4_JJ_ansatz}

In this section we study the 1/4 BPS states with $U(1)_{t}$$\times
SO(4)\times SO(2)~$symmetry. The geometries corresponding to such BPS states
have been studied in e.g. \cite{Chong:2004ce} -\cite{Lunin:2008tf}, which is
a $R_{t}$$\times S^{3}\times S^{1}$ fibration over 4d K\"{a}hler base. In
addition, there is a direction $y$,~as the product of two radii of $S^{3}~$%
and $S^{1}$. As argue in \cite{Chen:2007du}, the $S^{3}$ and $S^{1}~$shrinks
smoothly on two types of droplet regions in the 4d base as the direction $y$
goes to zero, i.e. $y=0$.

We first discuss the ansatz for the 1/4 BPS configurations. These
backgrounds have an additional $S^{1}$ isometry compared with the 1/8 BPS
backgrounds, and have a ten-dimensional solution of the form (in the
conventions of \cite{Donos:2006iy},\cite{Chen:2007du},\cite{Lunin:2008tf}),
\begin{eqnarray}
ds_{{10}}^{2} &=&-h^{-2}(dt+\omega )^{2}+h^{2}((Z+\frac{1}{2})^{{-1}%
}2\partial _{i}\partial _{{\bar{j}}}Kdz^{i}d{\bar{z}}^{\bar{j}%
}+dy^{2})+y(e^{G}d\Omega _{3}^{2}+e{^{-G}}(d\psi +{\mathcal{A}})^{2}),
\notag \\
F_{5} &=&\left\{ -d{[}y^{2}e^{{2G}}(dt+\omega ){]}-y^{2}(d\omega +\eta
\mathcal{F})+2i\partial \bar{\partial}K\right\} \wedge d\Omega ^{3}+\mathrm{%
dual,}  \notag \\
h^{{-2}} &=&2y\cosh G,\qquad  \notag \\
Z &=&\frac{1}{2}\mathrm{\tanh }G=-\frac{1}{2}y\partial _{y}(\frac{1}{y}%
\partial _{y}K),\,  \notag \\
d\omega &=&\frac{i}{2}d[\frac{1}{y}\partial _{{y}}(\bar{\partial}-\partial
)K]=\frac{i}{y}(\partial _{i}\partial _{{\bar{j}}}\partial _{{y}}Kdz_{i}d%
\bar{z}_{\bar{j}}+\partial _{\bar{{i}}}Zd\bar{z}_{i}dy-\partial _{{i}%
}Zdz_{i}dy),  \notag \\
2\eta \mathcal{F} &=&-i\partial \bar{\partial}D.  \label{1-4_ansatz}
\end{eqnarray}

The $K=K(z_{i},\bar{z}_{i};y),$ where $i=1,2,~$is the K\"{a}hler potential
for the 4d base, which also varies with the $y$ direction. $D$ is an
auxiliary function, and can be set to a constant, if the fibration of the $%
S^{1}$\ is a direct product, i.e. $\mathcal{A}=0$.~The volume of the 4d base
is constrained by a Monge-Ampere equation (as well as an auxiliary condition
for auxiliary function $D$),
\begin{eqnarray}
&&\log \det h_{i\bar{j}}=\log (Z+\frac{1}{2})+n\eta \log y+\frac{1}{y}%
(2-n\eta )\partial _{y}K+D(z_{i},\bar{z}_{\bar{j}}),  \label{eqn_general} \\
&&(1+\ast _{4})\partial \bar{\partial}D=\frac{4}{y^{2}}(1-n\eta )\partial
\bar{\partial}K.  \label{eq:MA2}
\end{eqnarray}%
In other words,
\begin{equation}
\det \partial _{i}\partial _{\bar{j}}K=(Z+\frac{1}{2})y^{n\eta }e^{\frac{1}{y%
}(2-n\eta )\partial _{y}K}e^{D}.  \label{MA_general}
\end{equation}

We can consider two types of 1/4 BPS states with the above ansatz. One class
is the states with two R-charges $J_{1},J_{2}$, which will be discussed in
this section. Another class is the states with R-charge $J$ and an $AdS$
spin or an $SO(4)$ spin $S_{1}$, and will be discussed in section \ref%
{sec:1-4_SJ}.$~$The first case corresponds to that the $S^{3}$ in the ansatz
(\ref{1-4_ansatz}) is in the $AdS$ directions, while the second case
corresponds to that the $S^{3}$ in the ansatz (\ref{1-4_ansatz}) is in the $%
S^{5}$ directions.

For the first 1/4 BPS sector, the dual operator is of the schematic form
\begin{equation}
O\sim {{\prod_{i=1}^{m}}}\mathrm{tr}(Z{}^{n_{1i}}Y{}^{n_{2i}})
\end{equation}%
where $Z,Y$ are two complex scalars of $\mathcal{N}$=4 SYM. The BPS bound is
satisfied as
\begin{equation}
\Delta -J_{1}-J_{2}=0.
\end{equation}

\subsection{Small $y$ analysis}

\label{sec:1-4_JJ_small_y}

For simplicity of the discussion, we first consider the case when the $S^{1}$
is a direct product factor, so we can set $\mathcal{A}=0,\mathcal{F}=d%
\mathcal{A}=0,~$as well as $n\eta =1.~$We can also set $D$ as a constant in (%
\ref{MA_general}). The equation is
\begin{equation}
\det \partial _{i}\partial _{\bar{j}}K=(1-4y^{2}\partial _{y^{2}}^{2}K)\frac{%
y}{8\sqrt{e}}e^{2\partial _{y^{2}}K}  \label{MA__ne1}
\end{equation}%
where we used $e^{D}=\frac{1}{4\sqrt{e}}$, and take derivatives with
respective to the $y^{2}$.

In this subsection we analyze the equations from (\ref{MA__ne1}) as well as
equations of regularity conditions at small $y$. We denotes $I_{-}$ as the
region $S^{3}\rightarrow 0$,$~Z=-\frac{1}{2};~$and $I_{+}$ as the region $%
S^{1}\rightarrow 0$,$~Z=\frac{1}{2}.~$The small $y$ behavior of the K\"{a}%
hler potential near two different types of droplet regions are very
different, and we will analyze the small $y$ equations in each droplet
regions.

In the region that $S^{1}$ shrinks, we have

\begin{equation}
Z=\frac{1}{2},~~~~\mathrm{~i.e.}~~-\frac{1}{2}y\partial _{y}(\frac{1}{y}%
\partial _{y}K)|_{y=0}=\frac{1}{2}  \label{boundary_condition_plus}
\end{equation}%
where the 1/2 BPS function $Z$ is related to $K$ via $Z=-\frac{1}{2}%
y\partial _{y}(\frac{1}{y}\partial _{y}K),~$and we have $h^{2}(Z+\frac{1}{2}%
)^{{-1}}=O(1)~$in the $Z=\frac{1}{2}~$region. The leading terms in $K$ that
have $z_{i},\bar{z}_{i}$ dependence will go as $K_{0}(z_{i},\bar{z}%
_{i})+y^{2}K_{1}(z_{i},\bar{z}_{i})$. In this region one can expand the $K$
as

{\large
\begin{equation}
K=-\frac{1}{4}y^{2}\log (y^{2})+K_{0}(z_{i},\bar{z}_{i})+y^{2}K_{1}(z_{i},%
\bar{z}_{i})+(y^{2})^{2}K_{2}(z_{i},\bar{z}_{i})
\end{equation}%
}i.e. up to order $(y^{2})^{3}.~$We also have the expansion of $Z=\frac{1}{2}%
-4y^{2}K_{2}.$

In the region that $S^{3}$ shrinks, we have $Z=-\frac{1}{2},~$and $h^{-2}(Z+%
\frac{1}{2})=o(y^{2}),$ so from the metric we require $\partial _{i}\partial
_{\bar{j}}K|_{y=0}=0$. Thereby in this region,
\begin{equation}
Z=-\frac{1}{2},~~~~~\mathrm{i.e.}~~-\frac{1}{2}y\partial _{y}(\frac{1}{y}%
\partial _{y}K)|_{y=0}=-\frac{1}{2},~~\ \mathrm{and~~\ }\partial
_{i}\partial _{\bar{j}}K|_{y=0}=0.  \label{boundary_condition_minus}
\end{equation}%
One can expand the $K$ in this region as

\begin{equation}
K=\frac{1}{4}y^{2}\log (y^{2})+K_{0}(z_{i},\bar{z}_{i})+y^{2}K_{1}(z_{i},%
\bar{z}_{i})+(y^{2})^{2}K_{2}(z_{i},\bar{z}_{i}),
\end{equation}%
up to order $(y^{2})^{3}$, with additional equation $\partial _{i}\partial _{%
\bar{j}}K_{0}=0$, due to $\partial _{i}\partial _{\bar{j}}K|_{y=0}=0$. A
necessary but weaker condition is $\det \partial _{i}\partial _{\bar{j}%
}K_{0}=0$\cite{Chen:2007du}. We also have the expansion of $Z=-\frac{1}{2}%
-4y^{2}K_{2}.$

We then analyze the equations for $K_{0},K_{1},K_{2}~$in those two regions
at $y$=0.

While in$~I_{-}$ i.e. $S^{3}\rightarrow 0$,$~Z=-\frac{1}{2}$, from (\ref%
{MA__ne1}),
\begin{eqnarray}
&&\det \partial _{i}\partial _{\bar{j}}K_{0}=0,  \label{small_y_minus_K0} \\
&&\partial _{i}\partial _{\bar{j}}K_{1}\partial _{k}\partial _{\bar{l}%
}K_{0}(\delta ^{i\bar{j}}\delta ^{k\bar{l}}-\delta ^{i\bar{l}}\delta ^{k\bar{%
j}})=0.  \label{small_y_minus_K1}
\end{eqnarray}%
In addition, from the regularity condition
\begin{equation}
\partial _{i}\partial _{\bar{j}}K_{0}=0.  \label{K0_minus}
\end{equation}%
These equations can be simultaneously solved by $\partial _{i}\partial _{%
\bar{j}}K_{0}=0$.

Assuming $\partial _{i}\partial _{\bar{j}}K_{0}$=0, which is a stronger
condition than $\det \partial _{i}\partial _{\bar{j}}K_{0}=0$, then from (%
\ref{MA__ne1})
\begin{equation}
\det \partial _{i}\partial _{\bar{j}}K_{1}+K_{2}e^{2K_{1}}=0.
\label{small_y_minus_K2}
\end{equation}%
This suggests that $K_{2}$ can be determined once knowing the expression of $%
K_{1}$, i.e.
\begin{equation}
K_{2}=-e^{-2K_{1}}\det \partial _{i}\partial _{\bar{j}}K_{1}.
\label{K2_minus}
\end{equation}

\vspace{1pt}

While in $I_{+}$ i.e. $S^{1}\rightarrow 0$,$~Z=\frac{1}{2}$, we get from (%
\ref{MA__ne1}),
\begin{eqnarray}
&&\det \partial _{i}\partial _{\bar{j}}K_{0}=\frac{1}{4e}e^{2K_{1}},
\label{small_y_plus_K0} \\
&&\partial _{i}\partial _{\bar{j}}K_{1}\partial _{k}\partial _{\bar{l}%
}K_{0}(\delta ^{i\bar{j}}\delta ^{k\bar{l}}-\delta ^{i\bar{l}}\delta ^{k\bar{%
j}})=0.  \label{small_y_plus_K1}
\end{eqnarray}%
The first equation can be written as%
\begin{equation}
K_{1}=\frac{1}{2}\log \det \partial _{i}\partial _{\bar{j}}K_{0}+\log 2+%
\frac{1}{2}.
\end{equation}

To summarize a little, in the $I_{-}~$region, we have equations (\ref%
{K0_minus}),(\ref{small_y_minus_K1}) for $K_{0},K_{1}$, and (\ref{K2_minus})
serves as the solution for $K_{2}.$ While in the $I_{+}~$region, we have the
coupled equations (\ref{small_y_plus_K0}),(\ref{small_y_plus_K1}) for $%
K_{0},K_{1}$ which can be solved.

One can also argue that in the limit that $I_{-}~$regions have zero measure
in terms of the 4d volume, the equations (\ref{small_y_plus_K0}),(\ref%
{small_y_plus_K1}) are valid in all the regions in 4d, except the loci of $%
I_{-}.~$

We now study how to solve these equations. We first study the $I_{+}~$%
region. We can expand in the large $r^{2}=$ $\left\vert z_{1}\right\vert
^{2}+\left\vert z_{2}\right\vert ^{2}~$region as
\begin{equation}
K_{0}=\frac{1}{2}(\left\vert z_{1}\right\vert ^{2}+\left\vert
z_{2}\right\vert ^{2})+\widetilde{K}_{0},\qquad K_{1}=\frac{1}{2}+\widetilde{%
K}_{1}.
\end{equation}%
From the leading order terms in the tilded variables we have
\begin{eqnarray}
&&\partial _{1}\partial _{\bar{1}}\widetilde{K}_{0}+\partial _{2}\partial _{%
\bar{2}}\widetilde{K}_{0}=\widetilde{K}_{1},  \label{K_tilde_eqn_01} \\
&&\partial _{1}\partial _{\bar{1}}\widetilde{K}_{1}+\partial _{2}\partial _{%
\bar{2}}\widetilde{K}_{1}=0.
\end{eqnarray}%
The expression of the $\widetilde{K}_{0}$ can always be obtained from $%
\widetilde{K}_{1}$ using the Green's function technique, due to equation (%
\ref{K_tilde_eqn_01}).

So we get in leading orders in large $\left\vert z_{1}\right\vert
^{2}+\left\vert z_{2}\right\vert ^{2}~$region, using 4d Green's function
method,
\begin{eqnarray}
K_{0} &=&\frac{1}{2}(\left\vert z_{1}\right\vert ^{2}+\left\vert
z_{2}\right\vert ^{2})-\frac{1}{2}\int_{{\normalcolor\mathcal{D}}}\mathrm{%
\log }(\left\vert z_{1}-z_{1}^{\prime }\right\vert ^{2}+\left\vert
z_{2}-z_{2}^{\prime }\right\vert ^{2})u(z_{i}^{\prime },\bar{z}_{i}^{\prime
})\frac{1}{n}d^{2}z_{1}^{\prime }d^{2}z_{2}^{\prime },  \label{K0_repulsion}
\\
K_{1} &=&\frac{1}{2}-\frac{1}{2}\int_{{\normalcolor\mathcal{D}}}\frac{1}{%
(\left\vert z_{1}-z_{1}^{\prime }\right\vert ^{2}+\left\vert
z_{2}-z_{2}^{\prime }\right\vert ^{2})}u(z_{i}^{\prime },\bar{z}_{i}^{\prime
})\frac{1}{n}d^{2}z_{1}^{\prime }d^{2}z_{2}^{\prime },
\end{eqnarray}%
where $\int_{{\normalcolor\mathcal{D}}}u(z_{i}^{\prime },\bar{z}_{i}^{\prime
})\frac{1}{n}d^{2}z_{1}^{\prime }d^{2}z_{2}^{\prime }=1$.$~$For convenience
we use a notation that $d^{2}z_{1}^{\prime }=\frac{1}{2i}dz_{1}^{\prime }d%
\bar{z}_{1}^{\prime }=dx^{\prime }dy^{\prime }$, and $(z_{i}^{\prime },\bar{z%
}_{i}^{\prime })$ denotes $(z_{1}^{\prime },\bar{z}_{1}^{\prime
},z_{2}^{\prime },\bar{z}_{2}^{\prime }).~$The logarithmic terms in $K_{0}$
in (\ref{K0_repulsion}) is very reminiscent of the repulsions between
eigenvalues in the droplet space.

Now we alternatively expand around a region near a particular radial
position, e.g. near $\left\vert z_{1}\right\vert ^{2}+\left\vert
z_{2}\right\vert ^{2}\simeq 1$, as
\begin{equation}
K_{0}=\frac{1}{2}+\frac{1}{4}(\left\vert z_{1}\right\vert ^{2}+\left\vert
z_{2}\right\vert ^{2}-1)^{2}+\widetilde{K}_{0},\qquad K_{1}=\frac{1}{2}\log
(\left\vert z_{1}\right\vert ^{2}+\left\vert z_{2}\right\vert ^{2}-1)+\frac{1%
}{2}+\widetilde{K}_{1}.
\end{equation}%
After taking $\left\vert z_{1}\right\vert ^{2}+\left\vert z_{2}\right\vert
^{2}\rightarrow 1$ limit, we get the leading order equations
\begin{eqnarray}
&&\left\vert z_{2}\right\vert ^{2}\partial _{1}\partial _{\bar{1}}\widetilde{%
K}_{0}+\left\vert z_{1}\right\vert ^{2}\partial _{2}\partial _{\bar{2}}%
\widetilde{K}_{0}-z_{1}\bar{z}_{2}\partial _{1}\partial _{\bar{2}}\widetilde{%
K}_{0}-z_{2}\bar{z}_{1}\partial _{2}\partial _{\bar{1}}\widetilde{K}_{0}=0,
\\
&&\left\vert z_{2}\right\vert ^{2}\partial _{1}\partial _{\bar{1}}\widetilde{%
K}_{1}+\left\vert z_{1}\right\vert ^{2}\partial _{2}\partial _{\bar{2}}%
\widetilde{K}_{1}-z_{1}\bar{z}_{2}\partial _{1}\partial _{\bar{2}}\widetilde{%
K}_{1}-z_{2}\bar{z}_{1}\partial _{2}\partial _{\bar{1}}\widetilde{K}_{1}=0.
\end{eqnarray}%
We see that $\widetilde{K}_{1}=-\frac{1}{2}(\left\vert z_{1}\right\vert
^{2}+\left\vert z_{2}\right\vert ^{2}-1)$ is an exact solution to the
equation, and $\widetilde{K}_{0}\simeq -\frac{1}{6}(\left\vert
z_{1}\right\vert ^{2}+\left\vert z_{2}\right\vert ^{2}-1)^{3}$ is a leading
order solution.

The equation is not exactly superposable, but in the small $\left\vert
z_{1}^{\prime }\right\vert ^{2}+\left\vert z_{2}^{\prime }\right\vert
^{2}\ll 1~$limit, it is approximately superposable, so in that limit,%
\begin{equation}
K_{1}\simeq \frac{1}{2}\log (\left\vert z_{1}\right\vert ^{2}+\left\vert
z_{2}\right\vert ^{2}-1)+1-\frac{1}{2}\int_{{\normalcolor\mathcal{D}}%
}(\left\vert z_{1}-z_{1}^{\prime }\right\vert ^{2}+\left\vert
z_{2}-z_{2}^{\prime }\right\vert ^{2})u(z_{i}^{\prime },\bar{z}_{i}^{\prime
})\frac{1}{n}d^{2}z_{1}^{\prime }d^{2}z_{2}^{\prime }
\end{equation}%
where $\int_{{\normalcolor\mathcal{D}}}u(z_{i}^{\prime },\bar{z}_{i}^{\prime
})\frac{1}{n}d^{2}z_{1}^{\prime }d^{2}z_{2}^{\prime }=1.~$This means that
there are extra small $Z=-\frac{1}{2}$ droplets located at ($z_{i}^{\prime },%
\bar{z}_{i}^{\prime }$), which are close to the origin and far from $%
\left\vert z_{1}\right\vert ^{2}+\left\vert z_{2}\right\vert ^{2}$=1, in
which case the superposition is possible.

\subsection{Small variation}

In this subsection we describe a change of variable that transforms (\ref%
{MA__ne1}) into linear equation, under the approximation that the changed
variable is slowly varying.

We consider the change of variable
\begin{equation}
K(z_{i},\bar{z}_{i},y)=\frac{1}{2}y^{2}-\frac{1}{4}y^{2}\log y^{2}+\frac{1}{2%
}(\left\vert z_{1}\right\vert ^{2}+\left\vert z_{2}\right\vert ^{2})+V(z_{i},%
\bar{z}_{i},y)  \label{change_variable_JJ}
\end{equation}%
and $V(z_{i},\bar{z}_{i},y)$ is a new function.

We have
\begin{equation}
\partial _{i}\partial _{\bar{j}}K=\frac{1}{2}\delta _{i\bar{j}}{}+\partial
_{i}\partial _{\bar{j}}V.
\end{equation}%
Then (\ref{MA__ne1}) becomes%
\begin{equation}
\frac{1}{4}+\frac{1}{2}\partial _{1}\partial _{\bar{1}}V+\frac{1}{2}\partial
_{2}\partial _{\bar{2}}V+(\partial _{1}\partial _{\bar{1}}V\partial
_{2}\partial _{\bar{2}}V-\partial _{1}\partial _{\bar{2}}V\partial
_{2}\partial _{\bar{1}}V)=\frac{1}{4}e^{2\partial
_{y^{2}}V}(1-2y^{2}\partial _{y^{2}}^{2}V).
\end{equation}%
$\allowbreak $Now if we assume slowly varying $V$
\begin{equation}
\partial _{i}V,\partial _{\bar{j}}V,\partial _{y^{2}}V\ll 1,
\end{equation}%
we then get
\begin{equation}
4(\partial _{1}\partial _{\bar{1}}V+\partial _{2}\partial _{\bar{2}%
}V)+y^{3}\partial _{y}(\frac{1}{y^{3}}\partial _{y}V)=0.
\label{V_eqn_linear_JJ}
\end{equation}

One can see that $V=-\frac{1}{2}\log (\left\vert z_{1}\right\vert
^{2}+\left\vert z_{2}\right\vert ^{2}+y^{2})$ are $V=\frac{y^{4}}{%
(\left\vert z_{1}\right\vert ^{2}+\left\vert z_{2}\right\vert ^{2}+y^{2})^{4}%
}$ are solutions to this linear and superposable equation.

One way to treat the equation is change of variable%
\begin{eqnarray}
&&\Psi =\frac{V}{y^{2}}, \\
&&4(\partial _{1}\partial _{\bar{1}}\Psi +\partial _{2}\partial _{\bar{2}%
}\Psi )+\frac{1}{y}\partial _{y}(y\partial _{y}\Psi )=\frac{4}{y^{2}}\Psi .
\label{poisson}
\end{eqnarray}%
The equation (\ref{poisson}) is a Poisson equation in 6d.

The solutions to (\ref{V_eqn_linear_JJ}) can be written as

\begin{eqnarray}  \label{V_solution_JJ}
V =\int_{{\normalcolor\mathcal{D}}}(-\frac{1}{2}\log (\left\vert
z_{1}-z_{1}^{\prime }\right\vert ^{2}+\left\vert z_{2}-z_{2}^{\prime
}\right\vert ^{2}+y^{2}))u(z_{i}^{\prime },\bar{z}_{i}^{\prime })\frac{1}{n}%
d^{2}z_{1}^{\prime }d^{2}z_{2}^{\prime }+\int_{{\normalcolor\mathcal{D}}}%
\frac{\alpha y^{4} u(z_{i}^{\prime },\bar{z}_{i}^{\prime })\frac{1}{n}%
d^{2}z_{1}^{\prime }d^{2}z_{2}^{\prime }}{(\left\vert z_{1}-z_{1}^{\prime
}\right\vert ^{2}+\left\vert z_{2}-z_{2}^{\prime }\right\vert ^{2}+y^{2})^{4}%
}  \notag \\
\end{eqnarray}%
where $\int_{{\normalcolor\mathcal{D}}}u(z_{i}^{\prime },\bar{z}_{i}^{\prime
})\frac{1}{n}d^{2}z_{1}^{\prime }d^{2}z_{2}^{\prime }=1.~$

The solution is valid in the region of slowly varying $V,$ which includes $%
y=0,Z=+\frac{1}{2},~$as one can check that (\ref{change_variable_JJ}), (\ref%
{V_solution_JJ}) satisfies the boundary condition (\ref%
{boundary_condition_plus}). It also includes large $y$ region$.$ The
solution breaks down near the $y=0,Z=-\frac{1}{2}$ droplets.

The Kahler potential can be written through (\ref{change_variable_JJ}). In
particular the Kahler potential at $y=0$ can be written as

\begin{eqnarray}
&&K|_{y=0}=\frac{1}{2}(\left\vert z_{1}\right\vert ^{2}+\left\vert
z_{2}\right\vert ^{2})+\int_{{\normalcolor\mathcal{D}}}\left( -\frac{1}{2}%
\log (\left\vert z_{1}-z_{1}^{\prime }\right\vert ^{2}+\left\vert
z_{2}-z_{2}^{\prime }\right\vert ^{2})\right) u(z_{i}^{\prime },\bar{z}%
_{i}^{\prime })\frac{1}{n}d^{2}z_{1}^{\prime }d^{2}z_{2}^{\prime },\;Z=+%
\frac{1}{2}  \notag \\
&&
\end{eqnarray}%
up to an overall constant shift. The$\;$$Z=-\frac{1}{2}$ droplets are like
conducting droplets in the 4d base.\vspace{1pt}

\subsection{Radially symmetric cases}

\label{sec:1-4_JJ_radial}

In this subsection, we look at the special case that $K=K(r^{2},y^{2})$,
which means that the K\"{a}hler potential only depends on the radial
direction $r^{2}=$ $\left\vert z_{1}\right\vert ^{2}+\left\vert
z_{2}\right\vert ^{2}~$of the 4d base and $y$.

The equation from (\ref{MA__ne1}) is (where we have introduced $e^{D}=\frac{1%
}{4\sqrt{e}}$)%
\begin{equation}
\partial _{r^{2}}K\partial _{r^{2}}(r^{2}\partial
_{r^{2}}K)=(1-4y^{2}\partial _{y^{2}}^{2}K)\frac{y}{8\sqrt{e}}e^{2\partial
_{y^{2}}K}  \label{K_eqn_radial}
\end{equation}%
where the derivatives are taken with respective to $r^{2}$ and $y^{2}.~$

We try to look for general solutions that $S^{3}$ shrinks at several
intervals in the $r$ direction, i.e.
\begin{equation}
I_{-}=(0,r_{1})\cup (r_{2},r_{3})\cup ...\cup (r_{2m},r_{2m+1})
\label{boundary_minus}
\end{equation}%
with $Z=-\frac{1}{2}$; while $S^{1}$ shrinks at
\begin{equation}
I_{+}=(r_{1},r_{2})\cup (r_{3},r_{4})\cup ...\cup (r_{2m+1},\infty )
\label{boundary_plus}
\end{equation}%
with $Z=\frac{1}{2}.$ Each region in $I_{-}$ (except the first) is a `shell'
of $Z=-\frac{1}{2}~$droplet. Each region in $I_{+}$ (except the last) is a
`shell' of $Z=\frac{1}{2}$ droplet. \ The superposable solutions in small $y$
in subsection \ref{sec:1-4_JJ_small_y} indicates that the above
configurations likely exist.

The $AdS_{5}\times S^{5}$ is an exact solution to this equation (\ref%
{K_eqn_radial}) as follows \cite{Chen:2007du}%
\begin{eqnarray}
K_{AdS} &=&\frac{1}{2}\left( \frac{1}{2}(r^{2}+y^{2}+1)+\sqrt{\frac{1}{4}%
(r^{2}+y^{2}-1)^{2}+y^{2}}\right)  \notag \\
&&-\frac{1}{2}\log \left( \frac{1}{2}(r^{2}+y^{2}+1)+\sqrt{\frac{1}{4}%
(r^{2}+y^{2}-1)^{2}+y^{2}}\right)  \notag \\
&&-\frac{1}{2}y^{2}\log \left( \frac{1}{2}(-r^{2}+y^{2}+1)+\sqrt{\frac{1}{4}%
(r^{2}+y^{2}-1)^{2}+y^{2}}\right) +\frac{1}{4}y^{2}\log y^{2}
\label{eq:kpot}
\end{eqnarray}%
and the $Z$ is
\begin{equation}
Z=\frac{1}{2}\frac{r^{2}+y^{2}-1}{\sqrt{(r^{2}+y^{2}-1)^{2}+4y^{2}}}
\label{Z_JJ}
\end{equation}%
where the variables are written in the unit that the $AdS$ radius $L=1$.~For
$AdS$, $I_{-}=(0,1)$ and $I_{+}=(1,\infty ).$

To get more intuition, we first look at the small $y$ expansion of the $%
AdS_{5}\times S^{5}~$solution:

When $S^{3}\rightarrow 0$, $Z=-\frac{1}{2}$,~i.e. $r^{2}<1$,
\begin{eqnarray}
&&K=\frac{1}{4}y^{2}\mathrm{\log }%
y^{2}+K_{0}+y^{2}K_{1}+(y^{2})^{2}K_{2}+o((y^{2})^{3}),\;
\label{K_S3_shrink} \\
&&K_{0}=\frac{1}{2},\qquad K_{1}=-\frac{1}{2}\log (1-r^{2}),\qquad K_{2}=-%
\frac{1}{4(r^{2}-1)^{2}}.
\end{eqnarray}

On the other hand, when $S^{1}\rightarrow 0$, $Z=\frac{1}{2}$, i.e. $r^{2}>1$%
,
\begin{eqnarray}
&&K=-\frac{1}{4}y^{2}\mathrm{\log }%
y^{2}+K_{0}+y^{2}K_{1}+(y^{2})^{2}K_{2}+o((y^{2})^{3}),\;
\label{K_S1_shrink} \\
&&K_{0}=\frac{1}{2}(r^{2}-\mathrm{\log }r^{2}),~K_{1}=\frac{1}{2}(1+\mathrm{%
\log }(r^{2}-1)-\mathrm{\log }r^{2}),~K_{2}=\frac{1}{4(r^{2}-1)^{2}}.
\label{K_S1_shrink_02}
\end{eqnarray}

The small $y$ equations for$~K$ with radial symmetry are:

When $r\in I_{-}$ i.e. $S^{3}\rightarrow 0$,$~Z=-\frac{1}{2}$,
\begin{eqnarray}
&&\partial _{r^{2}}K_{0}(\partial _{r^{2}}K_{0}+r^{2}\partial
_{r^{2}}^{2}K_{0})=0,  \label{I_minus_eqn_radial_1} \\
&&\partial _{r^{2}}K_{0}(\partial _{r^{2}}K_{1}+r^{2}\partial
_{r^{2}}^{2}K_{1})+\partial _{r^{2}}K_{1}(\partial
_{r^{2}}K_{0}+r^{2}\partial _{r^{2}}^{2}K_{0})=0,
\label{I_minus_eqn_radial_2} \\
&&\partial _{r^{2}}K_{0}(\partial _{r^{2}}K_{2}+r^{2}\partial
_{r^{2}}^{2}K_{2})+\partial _{r^{2}}K_{1}(\partial
_{r^{2}}K_{1}+r^{2}\partial _{r^{2}}^{2}K_{1})+K_{2}e^{2K_{1}}=0.
\end{eqnarray}

When $r\in I_{+}$ i.e. $S^{1}\rightarrow 0$,$~Z=\frac{1}{2}$,
\begin{eqnarray}
&&\partial _{r^{2}}K_{0}(\partial _{r^{2}}K_{0}+r^{2}\partial
_{r^{2}}^{2}K_{0})-\frac{1}{4e}e^{2K_{1}}=0,  \label{I_plus_eqn_radial_1} \\
&&\partial _{r^{2}}K_{0}(\partial _{r^{2}}K_{1}+r^{2}\partial
_{r^{2}}^{2}K_{1})+\partial _{r^{2}}K_{1}(\partial
_{r^{2}}K_{0}+r^{2}\partial _{r^{2}}^{2}K_{0})=0.
\label{I_plus_eqn_radial_2}
\end{eqnarray}%
We see that the AdS expression is one solution to the equations.

We first look at the coupled equations for $K_{0},K_{1}$ in region $r\in
I_{+}$ : \
\begin{eqnarray}
&&(K_{0}^{\prime })^{2}=-\frac{1}{8e}(e^{2K_{1}})^{\prime }\frac{1}{%
K_{1}^{\prime }+r^{2}K_{1}^{\prime \prime }},  \label{eq:-3} \\
&&r^{2}K_{0}^{\prime }K_{0}^{\prime \prime }=\frac{1}{4e}(e^{2K_{1}}+\frac{1%
}{2}(e^{2K_{1}})^{\prime }\frac{1}{K_{1}^{\prime }+r^{2}K_{1}^{\prime \prime
}}).
\end{eqnarray}%
where the primes denote $\partial _{r^{2}}$.$~$Dividing the 2nd equation
above by the 1st,
\begin{equation}
\frac{K_{0}^{\prime \prime }}{K_{0}^{\prime }}+\frac{K_{1}^{\prime \prime }}{%
K_{1}^{\prime }}=-\frac{2}{r^{2}}
\end{equation}%
which means
\begin{equation}
K_{1}^{\prime }=\frac{\gamma }{r^{4}K_{0}^{\prime }}.
\end{equation}%
This equation is satisfied for $AdS$ with $\gamma =\frac{1}{4}$. Plugging
this back into the 1st equation, we have
\begin{equation}
e^{-2K_{1}}(K_{1}^{\prime }+r^{2}K_{1}^{\prime \prime })+\frac{1}{4e\gamma
^{2}}r^{8}(K_{1}^{\prime })^{3}=0.
\end{equation}

If change of variable, $\frac{1}{2\sqrt{e}\gamma }e^{K_{1}}=q$,~the equation
is%
\begin{equation}
r^{2}qq^{\prime \prime }+qq^{\prime }+r^{8}q(q^{\prime
})^{3}-r^{2}(q^{\prime })^{2}=0.
\end{equation}%
For $AdS,~$we have $q=2(1-\frac{1}{r^{2}})^{\frac{1}{2}},$ where $r^{2}>1,~$%
and $\gamma =\frac{1}{4}.$

Now we look at scaling solutions near a particular $r_{\ast }^{2}$ region

\begin{equation}
p=qr_{\ast }^{2},\quad \;r^{2}=r_{\ast }^{2}(1+x),  \label{scaling_define}
\end{equation}%
where $\left\vert x\right\vert \ll 1$. $r_{\ast }^{2}$ can still be small
compared to the region near the boundary of the droplet space. The equation
becomes,$\quad $%
\begin{equation}
pp^{\prime \prime }+pp^{\prime }+p(p^{\prime })^{3}-(p^{\prime })^{2}=0
\end{equation}%
in which prime means $\partial _{x}$, with the inverted equation%
\begin{equation}
\ddot{x}-\dot{x}^{2}+\frac{1}{p}\dot{x}-1=0
\end{equation}%
where the dot denotes $\partial _{p}$.

From the last equation we get a class of solution

\begin{eqnarray}
&&r^{2}=r_{\ast }^{2}\int_{q_{-}r_{\ast }^{2}}^{qr_{\ast }^{2}}\frac{%
c_{1}J_{1}(p)+c_{2}Y_{1}(p)}{c_{1}J_{0}(p)+c_{2}Y_{0}(p)}dp+r_{-}^{2}, \\
&&r^{2}=-r_{\ast }^{2}\int_{qr_{\ast }^{2}}^{q_{+}r_{\ast }^{2}}\frac{%
c_{1}J_{1}(p)+c_{2}Y_{1}(p)}{c_{1}J_{0}(p)+c_{2}Y_{0}(p)}dp+r_{+}^{2},
\end{eqnarray}%
where $q_{-}$ is the value of $q$ at $r^{2}=r_{\ast }^{2}(1-x_{-})$ and $%
q_{+}$ is the value of $q$ at $r^{2}=r_{\ast }^{2}(1+x_{+})$, and $c_{1}$,$%
~c_{2}~$are constants. $J_{n}(p),Y_{n}(p)~$denote Bessel function of the
first kind and of the second kind, respectively.

So we have
\begin{eqnarray}
&&K_{1}^{\prime }=\frac{1}{r_{\ast }^{2}}\frac{c_{1}J_{0}(p)+c_{2}Y_{0}(p)}{%
c_{1}J_{1}(p)+c_{2}Y_{1}(p)}\frac{1}{p}, \\
&&K_{0}^{\prime }=\frac{\gamma }{r_{\ast }^{2}}\frac{%
c_{1}J_{1}(p)+c_{2}Y_{1}(p)}{c_{1}J_{0}(p)+c_{2}Y_{0}(p)}p,
\end{eqnarray}%
where the primes denote $\partial _{r^{2}}.~$Then,
\begin{eqnarray}
&&K_{0}=\int_{q_{-}r_{\ast }^{2}}^{qr_{\ast }^{2}}\gamma (\frac{%
c_{1}J_{1}(p)+c_{2}Y_{1}(p)}{c_{1}J_{0}(p)+c_{2}Y_{0}(p)})^{2}pdp+c_{-}, \\
&&K_{0}=-\int_{qr_{\ast }^{2}}^{q_{+}r_{\ast }^{2}}\gamma (\frac{%
c_{1}J_{1}(p)+c_{2}Y_{1}(p)}{c_{1}J_{0}(p)+c_{2}Y_{0}(p)})^{2}pdp+c_{+}, \\
&&K_{1}=\log (q)+\frac{1}{2}+\log (2\gamma ).
\end{eqnarray}

This solution is for a particular region near $r_{\ast }^{2}$,
i.e. $r_{\ast }^{2}(1-x_{-})\leqslant r^{2}\leqslant
r_{\ast }^{2}(1+x_{+}),$ due to the scaling limit taken in (\ref%
{scaling_define}).

\vspace{1pt}\vspace{1pt}Now we look at region $r\in I_{-}$ : In the range $%
r^{2}\in (r_{2k}^{2},~r_{2k+1}^{2})$, equation (\ref{I_minus_eqn_radial_1})
gives
\begin{equation}
K_{0}=c_{2k}+\alpha _{1}\log r^{2}  \label{eq:-2-1-1-2-3}
\end{equation}%
where $c_{2k},\alpha _{1}~$are constants in that interval. Then if $\alpha
_{1}\neq 0,$ (\ref{I_minus_eqn_radial_2}) would give $K_{1}=\alpha
_{2}+\alpha _{3}\log r^{2}$, which is quite different from the form of AdS,
near droplet boundary. So this suggests that $\alpha _{1}=0,~$and
\begin{equation}
K_{0}=c_{2k}  \label{eq:-2-1-1-2-3-2}
\end{equation}%
in the region $r^{2}\in (r_{2k}^{2},\,r_{2k+1}^{2}).$ This conclusion
applies rigorously in the case of radial symmetry.

\subsection{Non-radially symmetric solutions}

\label{sec:1-4_JJ_non_radial}

In this subsection we study solutions for the $K$ that are not radially
symmetric.

We look at the Kahler potential of the form
\begin{eqnarray}
K\text{$(z_{i},\bar{z}_{i},y)$} &\text{=}&\widetilde{K}%
(a^{2},y^{2})+y^{2}f(z_{i},\bar{z}_{i}),  \label{ansatz_non_radial} \\
a^{2} &=&a^{2}(z_{i},\bar{z}_{i})
\end{eqnarray}%
where $\widetilde{K}(a^{2},y^{2})~$is a regular solution to the radially
symmetric equation (\ref{K_eqn_radial})
\begin{equation}
\partial _{a^{2}}\widetilde{K}\partial _{a^{2}}(a^{2}\partial _{a^{2}}%
\widetilde{K})=(1-4y^{2}\partial _{y^{2}}^{2}\widetilde{K})\frac{y}{8\sqrt{e}%
}e^{2\partial _{y^{2}}\widetilde{K}}.  \label{radial_tilde}
\end{equation}%
Adding the $y^{2}f(z_{i},\bar{z}_{i})$ in (\ref{K_eqn_radial}) will not
change the regularity condition of the solution (\ref{K_eqn_radial}), since
\begin{equation}
K(z_{i},\bar{z}_{i},y)|_{y=0}=\widetilde{K}(a^{2}(z_{i},\bar{z}%
_{i}),y^{2})|_{y=0}.
\end{equation}%
This means that if $\widetilde{K}(a^{2},y^{2})~$is a solution that satisfies
the boundary conditions (\ref{boundary_condition_plus}),(\ref%
{boundary_condition_minus}), then $K(z_{i},\bar{z}_{i},y)$ also satisfies
the boundary conditions (\ref{boundary_condition_plus}),(\ref%
{boundary_condition_minus}). So the remaining work is to get the reduced
equations for $a^{2}(z_{i},\bar{z}_{i}),f(z_{i},\bar{z}_{i}),$ to guarantee
that $K(z_{i},\bar{z}_{i},y)$ is a solution to (\ref{MA__ne1}).

The role of $a^{2}(z_{i},\bar{z}_{i})$ is to describe the shape of the
droplets. The boundary between the two different droplets from the radially
symmetric solution is described by
\begin{equation}
a^{2}(z_{i},\bar{z}_{j})=\mathrm{const.}  \label{droplet_boundary}
\end{equation}%
where the constant could in principle take multiple values as discussed in
subsection \ref{sec:1-4_JJ_radial}.

For example, the $AdS$ solution is
\begin{equation}
a^{2}(z_{i},\bar{z}_{i})=|z_{1}|^{2}+|z_{2}|^{2},\quad \quad f(z_{i},\bar{z}%
_{i})=0.
\end{equation}%
So the boundary between two types of droplets is described by (\ref%
{droplet_boundary}), e.g. for $AdS$, as

\begin{equation}
a^{2}(z_{i},\bar{z}_{i})=|z_{1}|^{2}+|z_{2}|^{2}=1.
\end{equation}%
Changing the $a^{2}(z_{i},\bar{z}_{i})$ to a more general function thus
changes the droplet shape, while preserving regularity condition. The
resulting solution $K(z_{i},\bar{z}_{i},y)$ is no longer radially symmetric.

We then have
\begin{equation}
(1-4y^{2}\partial _{y^{2}}^{2}K)\frac{y}{8\sqrt{e}}e^{2\partial
_{y^{2}}K}=(1-4y^{2}\partial _{y^{2}}^{2}\widetilde{K})\frac{y}{8\sqrt{e}}%
e^{2\partial _{y^{2}}\widetilde{K}}e^{2f(z_{i},\bar{z}_{j})},  \label{rhs}
\end{equation}

\begin{equation}
\det \partial _{i}\partial _{\bar{j}}K\simeq a^{4}\partial _{i}\mathrm{\log }%
a^{2}\partial _{\bar{j}}\mathrm{\log }a^{2}\partial _{k}\partial _{\bar{l}}%
\mathrm{\log }a^{2}(\delta ^{i\bar{j}}\delta ^{k\bar{l}}-\delta ^{i\bar{l}%
}\delta ^{k\bar{j}})\partial _{a^{2}}\widetilde{K}\partial
_{a^{2}}(a^{2}\partial _{a^{2}}\widetilde{K})+\det (\partial _{i}\partial _{%
\bar{j}}\mathrm{\log }a^{2})(a^{2}\partial _{a^{2}}\widetilde{K})^{2},
\label{lhs}
\end{equation}%
where in the second equation (\ref{lhs}) we kept the leading order terms
pertaining to $\widetilde{K}$ in $\det \partial _{i}\partial _{\bar{j}}K$
above. So this is an approximation when $y^{2}f(z_{i},\bar{z}_{i})$ is much
smaller than $\widetilde{K}(a^{2},y^{2})~$in (\ref{ansatz_non_radial}), and
this is always correct in the small $y$ region.

Comparing (\ref{radial_tilde}) with (\ref{rhs}),(\ref{lhs}), we have
\begin{equation}
e^{2f(z_{i},\bar{z}_{j})}=a^{4}\partial _{i}\mathrm{\log }a^{2}\partial _{%
\bar{j}}\mathrm{\log }a^{2}\partial _{k}\partial _{\bar{l}}\mathrm{\log }%
a^{2}(\delta ^{i\bar{j}}\delta ^{k\bar{l}}-\delta ^{i\bar{l}}\delta ^{k\bar{j%
}})  \label{f_solution}
\end{equation}%
and%
\begin{equation}
\det (\partial _{i}\partial _{\bar{j}}\mathrm{\log }a^{2})=0.  \label{a_eqn}
\end{equation}

One can check that the $AdS$ expression satisfies the above equations. Once
we solve a solution for $a^{2}(z_{i},\bar{z}_{i})~$from (\ref{a_eqn}), then (%
\ref{f_solution}) already gives the solution for $f(z_{i},\bar{z}_{j}).$

We can solve the general function $a^{2}=a^{2}(z_{i},\bar{z}_{j})$ in two
steps. We define
\begin{equation}
\mathrm{\log }a^{2}=\mathrm{\log }\widetilde{a}^{2}-\mathrm{\log }(g^{2}).
\label{loga_JJ}
\end{equation}%
We look at a stronger condition than (\ref{a_eqn})
\begin{eqnarray}
&&\det (\partial _{i}\partial _{\bar{j}}\mathrm{\log }\widetilde{a}^{2})=0,
\label{loga_tilde} \\
&&(\delta ^{i\bar{l}}\delta ^{k\bar{j}}-\delta ^{i\bar{j}}\delta ^{k\bar{l}%
})\partial _{k}\partial _{\bar{l}}\mathrm{\log }\widetilde{a}^{2}\partial
_{i}\partial _{\bar{j}}\mathrm{\log }(g^{2})-\det (\partial _{i}\partial _{%
\bar{j}}\mathrm{\log }(g^{2}))=0.  \label{logg_general}
\end{eqnarray}%
These two equations add up to the equation (\ref{a_eqn}).

One way to treat equation (\ref{logg_general}) is to look at special
solutions satisfied by
\begin{equation}
\partial _{i}\partial _{\bar{j}}\mathrm{\log }(g^{2})=0.  \label{logg_hol}
\end{equation}%
One can look for special solutions%
\begin{equation}
\mathrm{\log }(g^{2})=\tilde{f}_{p}(p(z_{i},z_{j})+\bar{p}(\bar{z}_{i},\bar{z%
}_{j}))  \label{logg_special}
\end{equation}%
where $p$ is a function holomorphic in $z_{i},z_{j},~$and $\tilde{f}_{p}$ is
a parameter, e.g.
\begin{eqnarray}
&&p(z_{i},z_{j})=\sum_{l_{1},l_{2}\in Z^{+}}\epsilon
_{l_{1},l_{2}}(z_{1}^{l_{1}}z_{2}^{l_{2}}),  \label{ripple} \\
&&p(z_{i},z_{j})=\sum_{n_{1},n_{2}\in Z^{+}}\epsilon _{n_{1},n_{2}}(n_{1}%
\mathrm{\log }z_{1}+n_{2}\mathrm{\log }z_{2}),  \label{hole}
\end{eqnarray}%
etc, and these functions are superposable in (\ref{logg_special}).

We first let
\begin{equation}
\mathrm{\log }\widetilde{a}^{2}=\mathrm{\log }(|z_{1}|^{2}+|z_{2}|^{2})
\end{equation}%
which is a solution to (\ref{loga_tilde}), and the equation (\ref%
{logg_general}) becomes

\begin{equation}
\frac{1}{(|z_{1}|^{2}+|z_{2}|^{2})^{2}}z_{i}\bar{z}_{j}\partial _{i}\partial
_{\bar{j}}\log (g^{2})-\det (\partial _{i}\partial _{\bar{j}}\log (g^{2}))=0.
\label{g_eqn_02}
\end{equation}

There are several ways to treat this exact equation (\ref{g_eqn_02}). One
way is to consider $|\partial _{i}\partial _{\bar{j}}\log (g^{2})|\ll
|\partial _{i}\partial _{\bar{j}}\log \widetilde{a}^{2}|,$ then the equation
(\ref{g_eqn_02}) in the leading order reduces to%
\begin{equation}
z_{i}\bar{z}_{j}\partial _{i}\partial _{\bar{j}}\mathrm{\log }(g^{2})=0.
\end{equation}%
This equation is a Laplace equation in 4d in the variables $\log z_{i},\log
\bar{z}_{j}$.\ Another way to treat equation (\ref{logg_general}) is to look
at special solutions satisfied by $\partial _{i}\partial _{\bar{j}}\log
(g^{2})=0.$

It would also be nice to obtain more general solutions for (\ref{logg_hol}),
apart from (\ref{logg_special}). It is also possible to solve more general
cases in (\ref{logg_general}) when $\partial _{i}\partial _{\bar{j}}\log
(g^{2})\neq 0.$

Now we look at the solution
\begin{equation}
\mathrm{\log }a^{2}=\mathrm{\log }(|z_{1}|^{2}+|z_{2}|^{2})-\tilde{f}%
_{p}(p(z_{i},z_{j})+\bar{p}(\bar{z}_{i},\bar{z}_{j})).
\end{equation}

We denote $K_{y=0}=K_{d}$ in this section, where $d$ refers to the droplet
space. So we see that at $y=0$, $Z=\frac{1}{2},$
\begin{equation}
K_{d}=\frac{1}{2}a^{2}-\frac{1}{2}\mathrm{\log }a^{2}
\end{equation}%
up to an overall constant shift, where we used that $K_{y=0}=\widetilde{K}%
(a^{2},y^{2})|_{y=0}$, and also we used the $AdS~$expression for $\widetilde{%
K}$.

So we have%
\begin{eqnarray}
&&K_{d}=\frac{1}{2}(|z_{1}|^{2}+|z_{2}|^{2})-\frac{1}{2}\log (\left\vert
z_{1}\right\vert ^{2}+\left\vert z_{2}\right\vert ^{2})  \notag \\
&&-\frac{1}{2}(|z_{1}|^{2}+|z_{2}|^{2}-1)\tilde{f}_{p}(p(z_{i},z_{j})+\bar{p}%
(\bar{z}_{i},\bar{z}_{j})).
\end{eqnarray}%
For example,
\begin{eqnarray}
&&K_{d}=\frac{1}{2}(|z_{1}|^{2}+|z_{2}|^{2})-\frac{1}{2}\log (\left\vert
z_{1}\right\vert ^{2}+\left\vert z_{2}\right\vert ^{2})  \notag \\
&&-\frac{1}{2}(|z_{1}|^{2}+|z_{2}|^{2}-1)\tilde{f}_{p}(\sum_{l_{1},l_{2}\in
Z^{+}}\epsilon _{l_{1},l_{2}}(z_{1}^{l_{1}}z_{2}^{l_{2}})+c.c.)
\end{eqnarray}%
or
\begin{eqnarray}
&&K_{d}=\frac{1}{2}(|z_{1}|^{2}+|z_{2}|^{2})-\frac{1}{2}\log (\left\vert
z_{1}\right\vert ^{2}+\left\vert z_{2}\right\vert ^{2})  \notag \\
&&-\frac{1}{2}(|z_{1}|^{2}+|z_{2}|^{2}-1)\tilde{f}_{p}\sum_{n_{1},n_{2}\in
Z^{+}}\epsilon _{n_{1},n_{2}}(n_{1}\log \left\vert z_{1}\right\vert
^{2}+n_{2}\log \left\vert z_{2}\right\vert ^{2})
\end{eqnarray}%
etc.

In the case of (\ref{ripple}), it changes the droplet shape to, according to
(\ref{droplet_boundary}),
\begin{equation}
\left\vert z_{1}\right\vert ^{2}+\left\vert z_{2}\right\vert ^{2}=1+\tilde{f}%
_{p}(p(z_{i},z_{j})+\bar{p}(\bar{z}_{i},\bar{z}_{j}))=1+\tilde{f}%
_{p}(\sum_{l_{1},l_{2}\in Z^{+}}\epsilon
_{l_{1},l_{2}}(z_{1}^{l_{1}}z_{2}^{l_{2}})+c.c.)
\end{equation}%
where the $\epsilon $'s are small, so it may describe adding small ripples
to the solution corresponding to $\widetilde{a}^{2}.~l_{1},l_{2}$ may be
considered as the wave-numbers of the ripples. In the case of (\ref{hole}),
it may describe adding separate $Z=-\frac{1}{2}$ droplets in the 4d base.

To summarize a little, we considered $\log \widetilde{a}^{2}$ as a solution
before adding ripples, and $\log a^{2}$ as the solution adding ripples on $%
\log \widetilde{a}^{2}$. The droplet boundary is described by (\ref%
{droplet_boundary}).

We can also consider more generally
\begin{equation}
\mathrm{\log }\widetilde{a}^{2}=\log (\left\vert z_{1}-z_{1}^{\prime
}\right\vert ^{2}+\left\vert z_{2}-z_{2}^{\prime }\right\vert ^{2})
\end{equation}%
since%
\begin{equation}
\det (\partial _{i}\partial _{\bar{j}}\log (\left\vert z_{1}-z_{1}^{\prime
}\right\vert ^{2}+\left\vert z_{2}-z_{2}^{\prime }\right\vert ^{2}))=0.
\end{equation}

One can check
\begin{equation}
z_{i}\bar{z}_{j}\partial _{i}\partial _{\bar{j}}\log (g^{2})=0
\end{equation}%
has a special exact solution
\begin{equation}
\log (g^{2})=\frac{(\bar{c}_{1}z_{1}+\bar{c}_{2}z_{2})+c.c.}{%
|z_{1}|^{2}+|z_{2}|^{2}}
\end{equation}%
one can approximately superpose them and get the solution which is correct
in the leading order approximation in the large $z$ approximation%
\begin{equation}
\mathrm{\log }\widetilde{a}^{2}\simeq \frac{1}{n}\sum_{k}\log (\left\vert
z_{1}-z_{1,k}\right\vert ^{2}+\left\vert z_{2}-z_{2,k}\right\vert ^{2})
\end{equation}%
or%
\begin{equation}
\mathrm{\log }\widetilde{a}^{2}\simeq \int_{{\normalcolor\mathcal{D}}}\log
(\left\vert z_{1}-z_{1}^{\prime }\right\vert ^{2}+\left\vert
z_{2}-z_{2}^{\prime }\right\vert ^{2})\frac{u(z_{i}^{\prime },z_{\bar{\imath}%
}^{\prime })}{n}d^{2}z_{1}^{\prime }d^{2}z_{2}^{\prime }
\end{equation}%
where $\int_{{\normalcolor\mathcal{D}}}\frac{u}{n}d^{2}z_{1}^{\prime
}d^{2}z_{2}^{\prime }=1,~$and$~u(z_{i}^{\prime },z_{\bar{\imath}}^{\prime })$
is a density function.

Then
\begin{equation}
\sum_{k}\frac{1}{n}\frac{(z_{i}-z_{i,k})(\bar{z}_{j}-\bar{z}_{j,k})}{%
(\left\vert z_{1}-z_{1,k}\right\vert ^{2}+\left\vert
z_{2}-z_{2,k}\right\vert ^{2})^{2}}\partial _{i}\partial _{\bar{j}}\log
(g^{2})-\det (\partial _{i}\partial _{\bar{j}}\log (g^{2}))\simeq 0.
\end{equation}%
One can look for special solutions to the stronger equation $\partial
_{i}\partial _{\bar{j}}\log (g)=0$, e.g. (\ref{logg_special}).

We have then

\begin{equation}
\log a^{2}\simeq \sum_{k}\frac{_{1}}{n}\log (\left\vert
z_{1}-z_{1,k}\right\vert ^{2}+\left\vert z_{2}-z_{2,k}\right\vert ^{2})-%
\tilde{f}_{p}(p(z_{i},z_{j})+\bar{p}(\bar{z}_{i},\bar{z}_{j}))
\end{equation}%
or%
\begin{equation}
\log a^{2}\simeq \int_{{\normalcolor\mathcal{D}}}\log (\left\vert
z_{1}-z_{1}^{\prime }\right\vert ^{2}+\left\vert z_{2}-z_{2}^{\prime
}\right\vert ^{2})\frac{u}{n}d^{2}z_{1}^{\prime }d^{2}z_{2}^{\prime }-\tilde{%
f}_{p}(p(z_{i},z_{j})+\bar{p}(\bar{z}_{i},\bar{z}_{j})).
\end{equation}

So we see that at $y=0$,$~Z=\frac{1}{2},$ where we have denoted $%
K_{y=0}=K_{d}$ in this section,
\begin{eqnarray}
K_{d} &\simeq &\frac{1}{2}(|z_{1}|^{2}+|z_{2}|^{2})-\frac{1}{2}\sum_{k}\frac{%
_{1}}{n}\log (\left\vert z_{1}-z_{1,k}\right\vert ^{2}+\left\vert
z_{2}-z_{2,k}\right\vert ^{2})  \notag \\
&&-\frac{1}{2}(|z_{1}|^{2}+|z_{2}|^{2}-1)\tilde{f}_{p}(p(z_{i},z_{j})+\bar{p}%
(\bar{z}_{i},\bar{z}_{j})),
\end{eqnarray}%
e.g.
\begin{eqnarray}
K_{d} &\simeq &\frac{1}{2}(|z_{1}|^{2}+|z_{2}|^{2})-\frac{1}{2}\sum_{k}\frac{%
_{1}}{n}\log (\left\vert z_{1}-z_{1,k}\right\vert ^{2}+\left\vert
z_{2}-z_{2,k}\right\vert ^{2})  \notag \\
&&-\frac{1}{2}(|z_{1}|^{2}+|z_{2}|^{2}-1)\tilde{f}_{p}(\sum_{l_{1},l_{2}\in
Z^{+}}\epsilon _{l_{1},l_{2}}(z_{1}^{l_{1}}z_{2}^{l_{2}})+c.c.)
\end{eqnarray}%
or
\begin{eqnarray}
K_{d} &\simeq &\frac{1}{2}(|z_{1}|^{2}+|z_{2}|^{2})-\frac{1}{2}\sum_{k}\frac{%
_{1}}{n}\log (\left\vert z_{1}-z_{1,k}\right\vert ^{2}+\left\vert
z_{2}-z_{2,k}\right\vert ^{2})  \notag \\
&&-\frac{1}{2}(|z_{1}|^{2}+|z_{2}|^{2}-1)\tilde{f}_{p}(\sum_{n_{1},n_{2}\in
Z^{+}}\epsilon _{n_{1},n_{2}}(n_{1}\mathrm{log}z_{1}+n_{2}\mathrm{log}%
z_{2})+c.c.)
\end{eqnarray}%
etc; or, in the continuous approximation
\begin{eqnarray}
K_{d} &\simeq &\frac{1}{2}(|z_{1}|^{2}+|z_{2}|^{2})-\frac{1}{2}\int_{{%
\normalcolor\mathcal{D}}}\log (\left\vert z_{1}-z_{1}^{\prime }\right\vert
^{2}+\left\vert z_{2}-z_{2}^{\prime }\right\vert ^{2})\frac{u}{n}%
d^{2}z_{1}^{\prime }d^{2}z_{2}^{\prime }  \notag \\
&&-\frac{1}{2}(|z_{1}|^{2}+|z_{2}|^{2}-1)\tilde{f}_{p}(p(z_{i},z_{j})+\bar{p}%
(\bar{z}_{i},\bar{z}_{j})).
\end{eqnarray}

We argue that $K_{d}(z_{i},\bar{z}_{i})$ can be interpreted as the effective
Hamiltonian of a test eigenvalue in the dual $\mathcal{N}$=4 SYM. We may
also interpret $\vspace{1pt}-\partial _{i}K_{d}(z_{i},\bar{z}_{i})$ as the
force experienced by the test eigenvalue along the $i$-direction. For
example, for $AdS$ (where the $L$ is restored)

\begin{eqnarray}
&&K_{d}(z_{i},\bar{z}_{i})=\frac{1}{2}(\left\vert z_{1}\right\vert
^{2}+\left\vert z_{2}\right\vert ^{2})-\frac{1}{2}L^{2}\log (\left\vert
z_{1}\right\vert ^{2}+\left\vert z_{2}\right\vert ^{2}), \\
&&-\partial _{r}K_{d}(z_{i},\bar{z}_{i})=-r+\frac{L^{2}}{r},~~\quad
r\geqslant L.
\end{eqnarray}%
We see that the forces become zero at special $r=L.$

Now we consider the matrix model methods of \cite{Berenstein:2005aa},\cite%
{Correa:2010zj},\cite{Berenstein:2007wz},
\begin{equation}
\mathcal{H}=\frac{1}{2}\mathrm{tr}\left( \left\vert D_{0}\Phi
_{1}\right\vert ^{2}+\left\vert \Phi _{1}\right\vert ^{2}+\left\vert
D_{0}\Phi _{2}\right\vert ^{2}+\left\vert \Phi _{2}\right\vert ^{2}\right)
\end{equation}%
where these two matrices are the zero modes of two complex scalars $Z,Y~$in $%
\mathcal{N}$=4 SYM. The special case of one matrix has been studied in \cite%
{Corley:2001zk},\cite{Berenstein:2004kk},\cite{Lin:2004nb} and both the
gravity side and gauge side have been matched \cite{Corley:2001zk},\cite%
{Berenstein:2004kk},\cite{Lin:2004nb}. When reducing to the eigenvalue
basis, the wavefunction of eigenvalues acquires a factor from the measure
when integrating out off-diagonal components. The measure terms were derived
in \cite{Berenstein:2005aa} for multiple matrices at strong coupling. We can
also include interaction terms for multiple matrices and at large $N$ this
is relevant for the non-BPS states. Related issues are also discussed in
e.g. \cite{Brown:2007xh} -\cite{Bhattacharyya:2008rb}.

We can consider the wavefunction norm defining an effective Hamiltonian for
the eigenvalues as in \cite{Berenstein:2005aa}, i.e.
\begin{equation}
\left\langle \psi \mid \psi \right\rangle \sim e^{-2\mathcal{H}_{eff}}
\label{psi_norm_JJ}
\end{equation}%
where $\psi $ is a wavefunction for multiple eigenvalues. The wavefunction
norm also gives the probability density function of the eigenvalue
distributions.

The effective Hamiltonian from (\ref{psi_norm_JJ}) \cite{Berenstein:2005aa}
is
\begin{equation}
\mathcal{H}_{eff}=\frac{1}{2}{\sum_{k}}\left( \left\vert u_{1,k}\right\vert
^{2}+\left\vert u_{2,k}\right\vert ^{2}\right) -\frac{1}{2}\sum_{j<k}\log
(\left\vert u_{1,k}-u_{1,j}\right\vert ^{2}+\left\vert
u_{2,k}-u_{2,j}\right\vert ^{2})
\end{equation}%
where we used a different convention for $\mathcal{H}_{eff}$ in the exponent
in (\ref{psi_norm_JJ}) from the standard notation in \cite{Berenstein:2005aa}%
, to have the mass terms having canonical $\frac{1}{2}\left\vert
u\right\vert ^{2}~$forms, and there is no essential difference. We denote
the eigenvalues of the matrices or fields from the gauge theory as $u_{i,k},%
\bar{u}_{i,k}$, where $i$ labels the dimensions and $k$ (or $j$) labels the
individual eigenvalues.

We make an identification%
\begin{equation}
z_{i,k}=\frac{L}{\sqrt{N}}u_{i,k}
\end{equation}%
\begin{equation}
\frac{L^{2}}{N}\mathcal{H}_{eff}=\frac{1}{2}{\sum_{k}}\left( \left\vert
z_{1,k}\right\vert ^{2}+\left\vert z_{2,k}\right\vert ^{2}\right) -\frac{1}{2%
}L^{2}\sum_{j<k}\frac{1}{N}\log (\left\vert z_{1,k}-z_{1,j}\right\vert
^{2}+\left\vert z_{2,k}-z_{2,j}\right\vert ^{2})
\end{equation}%
up to a shift of constant $c$, and $c=-\frac{L^{2}(N-1)}{4}\log \frac{N}{%
L^{2}}$ in this case.

We denote the effective Hamiltonian for a test eigenvalue as $H_{eff}(u_{i},%
\bar{u}_{i})$, where $u_{i},\bar{u}_{i}$ denotes the test eigenvalue.$~$So
the effective Hamiltonian for a test eigenvalue, $H_{eff}(u_{i},\bar{u}%
_{i}), $ is given by, where the $L$ is restored,%
\begin{eqnarray}
\frac{L^{2}}{N}H_{eff} &=&\frac{1}{2}\left( \left\vert z_{1}\right\vert
^{2}+\left\vert z_{2}\right\vert ^{2}\right) -\frac{1}{2}L^{2}\sum_{j}\frac{1%
}{N}\log (\left\vert z_{1}-z_{1,j}\right\vert ^{2}+\left\vert
z_{2}-z_{2,j}\right\vert ^{2})  \notag \\
&\simeq &\frac{1}{2}\left( \left\vert z_{1}\right\vert ^{2}+\left\vert
z_{2}\right\vert ^{2}\right) -\frac{1}{2}L^{2}\log \left( \left\vert
z_{1}\right\vert ^{2}+\left\vert z_{2}\right\vert ^{2}\right) .
\label{K_d_JJ}
\end{eqnarray}%
The most probable eigenvalue density distribution has an $S^{3}~$symmetry,
and the last term is a sum over the eigenvalues, whose leading term is such
that the overall force is exerted from the origin. We have used the notation
$\mathcal{H}_{eff}$ to denote the effective Hamiltonian of a system of $N$
eigenvalues, while we have used $H_{eff}(u_{i},\bar{u}_{i})$ to denote the
effective Hamiltonian of one test eigenvalue $(u_{i},\bar{u}_{i})$.

So we have
\begin{equation}
K_{d}(\frac{L}{\sqrt{N}}u_{i},\frac{L}{\sqrt{N}}\bar{u}_{i})=\frac{L^{2}}{N}%
H_{eff}(u_{i},\bar{u}_{i})=-\frac{1}{2}\frac{L^{2}}{N}\log \left\langle \psi
_{b}(u_{i},\bar{u}_{i})\mid \psi _{b}(u_{i},\bar{u}_{i})\right\rangle
\end{equation}%
where $\frac{L}{\sqrt{N}}u_{i}=z_{i},$ and the $\psi _{b}(u_{i},\bar{u_{i}})$
denotes the reduced wavefunction for the test eigenvalue, and $H_{eff}$ here
is the effective Hamiltonian for the test eigenvalue. We have $i=1,2$ for
1/4 BPS case, in this section. However, this interpretation that we have
suggested here may not be the only interpretation.

For the excited states, the wavefunction changes, and thus the effective
Hamiltonian also changes and can be defined via e.g. (\ref{psi_norm_JJ})
\cite{Berenstein:2005aa}, \cite{Correa:2010zj}.

Finally, we suggest that the geometries corresponding to $\log a^{2},$ e.g.
in (\ref{loga_JJ}),(\ref{ripple}), would correspond to the operators of the
schematic form

\begin{equation}
O\sim {{\prod_{l_{1},l_{2}}}}e^{\mathrm{tr}(Z{}^{l_{1}}Y{}^{l_{2}})}O_{\chi }
\end{equation}%
where $O_{\chi }$, or the superposition thereof, may be the operator dual to
the geometry corresponding to $\log \widetilde{a}^{2}.$ More generally, $%
O_{\chi }$ may be related to solutions to (\ref{eqn_two_radial}).$~$Similar
1/4 BPS operators of the form related to $O_{\chi }$ have been analyzed in
e.g. \cite{Brown:2007xh} -\cite{deMelloKoch:2009jc}. The first factor can be
considered as the coherent states, whose expansions are superpositions of
polynomials of traces. The ripples can be regarded as the collective
phenomenon of many eigenvalues. The ripples here are analogous to the
ripples in the 1/2 BPS case, e.g. \cite{Berenstein:2005aa},\cite%
{Grant:2005qc},\cite{Takayama:2005yq}.\vspace{1pt}

\section{1/8 BPS geometries with $J_1, J_2, J_3$}

\label{sec:1-8_JJJ}

\subsection{General ansatz}

In this section we study the 1/8 BPS states with $U(1)_{t}$$\times SO(4)~$%
symmetry. The geometries corresponding to such BPS states have been studied
in \cite{Kim:2005ez} and in \cite{Chen:2007du},\cite{Gauntlett:2006ns},\cite%
{Gava:2006pu} (see also related discussion \cite{Gauntlett:2007ts}, \cite%
{Lunin:2008tf}), which is a $R_{t}$$\times S^{3}$ fibration over 6d K\"{a}%
hler base. As argued in \cite{Chen:2007du}, the $S^{3}$ can shrink smoothly
on the location of 5d surfaces in the 6d base.

In the conventions of \cite{Kim:2005ez}, and \cite{Chen:2007du}, we can
write the ten dimensional ansatz as
\begin{eqnarray}
ds_{10}^{2} &=&-e^{2\alpha }(dt+\omega )^{2}+e^{-2\alpha }(2\partial
_{i}\partial _{{\bar{j}}}K)dz^{i}d{\bar{z}}^{\bar{j}}+e^{2\alpha }d\Omega
_{3}^{2},  \notag \\
F_{5} &=&(d[e^{4\alpha }(dt+\omega )]-2i\eta \partial \bar{\partial}K)\wedge
\Omega _{3}+\mathrm{dual,}  \label{1-8_ansatz}
\end{eqnarray}%
where
\begin{eqnarray}
\qquad \ 2\eta d\omega &=&\mathcal{R},\mathcal{~}e^{-4\alpha }=-{\textstyle%
\frac{1}{8}}R,~~~  \notag  \label{eq:7fal} \\
\square _{6}e^{-4\alpha } &=&{\textstyle\frac{1}{8}}(R_{ab}R^{ab}-{\textstyle%
\frac{1}{2}}R^{2}).
\end{eqnarray}%
$R$ and $\mathcal{R}$ are the Ricci scalar and Ricci form of the 6d base $%
h_{ab}dx^{a}dx^{b}=2\partial _{i}\partial _{{\bar{j}}}Kdz^{i}d\bar{z}^{\bar{j%
}}$, where $\square _{6}$ is with respect to the metric $h_{ab}$, and $%
K=K(z_{i},\bar{z}_{i}),~i=1,2,3,$ is the K\"{a}hler potential of the 6d
base. The $y$ direction for the 1/8 BPS ansatz can be considered as $%
y^{2}=e^{2\alpha }.~$

One can also consider two types of 1/8 BPS states with the above ansatz. One
type is the states with three R-charges $J_{1},J_{2},J_{3},$ which will be
discussed in this section. Another type is the states with R-charge $J$ and $%
AdS$ spins or $SO(4)$ spins $S_{1},S_{2}$, which will be discussed in
section \ref{sec:1-8_SSJ}.$~$The first case corresponds to that the $S^{3}$
in the ansatz (\ref{1-8_ansatz}) is in the $AdS$ directions, while the
second case corresponds to that the $S^{3}$ in the ansatz (\ref{1-8_ansatz})
is in the $S^{5}$ directions.

For the first 1/8 BPS sector, the dual operator we consider is of the
schematic form
\begin{equation}
O\sim {{\prod_{i=1}^{m}}}\text{tr}(Z{}^{n_{1i}}Y{}^{n_{2i}}X{}^{n_{3i}})
\end{equation}%
where $Z,Y,X$ are three complex scalars of $\mathcal{N}$=4 SYM. The BPS
bound is satisfied as
\begin{equation}
\Delta -J_{1}-J_{2}-J_{3}=0.
\end{equation}

\subsection{AdS}

\label{sec:1-8_JJJ_AdS_eigen}

Let's first study the most symmetric case, the $AdS_{5}\times S^{5}.$ The K%
\"{a}hler potential for the $AdS_{5}\times S^{5}$ solution is \cite%
{Chen:2007du} (in the unit that the $AdS$ radius $L$=1)%
\begin{equation}
K=\frac{1}{2}(|z_{1}|^{2}+|z_{2}|^{2}+|z_{3}|^{2})-\frac{1}{2}\log
(|z_{1}|^{2}+|z_{2}|^{2}+|z_{3}|^{2})
\end{equation}%
i.e.%
\begin{eqnarray}
&&K=\frac{1}{2}|\vec{r}|^{2}-\log |\vec{r}-\vec{0}|, \\
&&-\nabla K=\frac{1}{|\vec{r}-0|}\vec{e}_{r}-\vec{r}  \label{force_AdS}
\end{eqnarray}%
where $\vec{r}~$is the radial vector in 6d. As similar to the discussion at
the end of section \ref{sec:1-4_JJ_non_radial}, we argue that $-\nabla K$
can be considered as an overall force experienced by the test eigenvalue
(which includes the forces exerted from all other eigenvalues). In the case
of $AdS$, the overall force is exerted as if from the origin, e.g. in (\ref%
{force_AdS}), and it can be interpreted that the eigenvalues are uniform in
angular directions. This enlarged symmetry is related to the $SO(6)$
symmetry of the $\mathcal{N}$=4 SYM.

One can also restore the $AdS$ radius $L$,%
\begin{eqnarray}
&&K=\frac{1}{2}(|z_{1}|^{2}+|z_{2}|^{2}+|z_{3}|^{2})-\frac{1}{2}L^{2}\log
(|z_{1}|^{2}+|z_{2}|^{2}+|z_{3}|^{2}), \\
&&r^{2}=\left\vert z_{1}\right\vert ^{2}+\left\vert z_{2}\right\vert
^{2}+\left\vert z_{3}\right\vert ^{2},  \notag \\
&&-\partial _{r}K(z_{i},\bar{z}_{i})=\frac{L^{2}}{r}-r.
\end{eqnarray}%
The force balance condition $-\partial _{r}K(z_{i},\bar{z}_{i})=0$ amounts to%
\begin{equation}
r=L.
\end{equation}

Now we consider the matrix model methods of \cite{Berenstein:2005aa}, \cite%
{Correa:2010zj}, \cite{Berenstein:2007wz},
\begin{equation}
\mathcal{H}=\frac{1}{2}\mathrm{tr}\left( \left\vert D_{0}\Phi
_{1}\right\vert ^{2}+\left\vert \Phi _{1}\right\vert ^{2}+\left\vert
D_{0}\Phi _{2}\right\vert ^{2}+\left\vert \Phi _{2}\right\vert
^{2}+\left\vert D_{0}\Phi _{3}\right\vert ^{2}+\left\vert \Phi
_{3}\right\vert ^{2}\right)
\end{equation}%
where these three matrices are the zero modes of three complex scalars $%
Z,Y,X $ in $\mathcal{N}$=4 SYM. The wavefunction norm defines the effective
Hamiltonian \cite{Berenstein:2005aa}

\begin{equation}
\left\langle \psi \mid \psi \right\rangle \sim e^{-2\mathcal{H}_{eff}},
\label{eq:7kgreens-1-1-1-1-1}
\end{equation}

\begin{equation}
\mathcal{H}_{eff}=\frac{1}{2}\sum_{k}\left( \left\vert u_{1,k}\right\vert
^{2}+\left\vert u_{2,k}\right\vert ^{2}+\left\vert u_{3,k}\right\vert
^{2}\right) -\frac{1}{2}\sum_{j<k}\log (\left\vert
u_{1,k}-u_{1,j}\right\vert ^{2}+\left\vert u_{2,k}-u_{2,j}\right\vert
^{2}+\left\vert u_{3,k}-u_{3,j}\right\vert ^{2}),
\end{equation}%
at strong coupling and large $N$, where the eigenvalues of the matrices or
fields from the gauge theory are denoted as $u_{i,k},\bar{u}_{i,k}$. We use $%
i$ to label dimensions, and $j,k$ to label eigenvalues.

Similar to section \ref{sec:1-4_JJ_non_radial}, we make an identification
\begin{equation}
z_{i,k}=\frac{L}{\sqrt{N}}u_{i,k}
\end{equation}%
\begin{eqnarray}
&&{{\frac{L^{2}}{N}}}\mathcal{H}{{_{eff}}}{\small =}{{\frac{1}{2}}}%
\sum_{k}(\left\vert z_{1,k}\right\vert ^{2}+\left\vert z_{2,k}\right\vert
^{2}+\left\vert z_{3,k}\right\vert ^{2}){\small -}{{\frac{1}{2}}{\small L}%
^{2}}\sum_{j<k}\textstyle{\frac{1}{N}\log {\small (}\left\vert
z_{1,k}-z_{1,j}\right\vert ^{2}{\small +}\left\vert
z_{2,k}-z_{2,j}\right\vert ^{2}{\small +}\left\vert
z_{3,k}-z_{3,j}\right\vert ^{2}{\small )}}  \notag  \label{eq:-18} \\
&&
\end{eqnarray}%
up to a shift of constant $c~$and $c=-\frac{L^{2}(N-1)}{4}\log \frac{N}{L^{2}%
}$ in this case.

So the effective Hamiltonian for a test eigenvalue $H_{eff}(u_{i},\bar{u}%
_{i})$ is, similar to the discussion in section \ref{sec:1-4_JJ_non_radial},
\begin{eqnarray}
\frac{L^{2}}{N}H_{eff} &=&\frac{1}{2}\left( \left\vert z_{1}\right\vert
^{2}+\left\vert z_{2}\right\vert ^{2}+\left\vert z_{3}\right\vert
^{2}\right) -\frac{1}{2}L^{2}\sum_{j}\frac{1}{N}\log (\left\vert
z_{1}-z_{1,j}\right\vert ^{2}+\left\vert z_{2}-z_{2,j}\right\vert
^{2}+\left\vert z_{1}-z_{3,j}\right\vert ^{2})  \notag \\
&\simeq &\frac{1}{2}\left( \left\vert z_{1}\right\vert ^{2}+\left\vert
z_{2}\right\vert ^{2}+\left\vert z_{3}\right\vert ^{2}\right) -\frac{1}{2}%
L^{2}\log \left( \left\vert z_{1}\right\vert ^{2}+\left\vert
z_{2}\right\vert ^{2}+\left\vert z_{3}\right\vert ^{2}\right) .
\label{K_d_JJJ}
\end{eqnarray}%
The most probable eigenvalue density distribution would have an $S^{5}$
symmetry, so the last term is a sum over the eigenvalues, whose net effect
in the leading order would be that the force would be exerted from the
origin.

So we see that
\begin{equation}
K_{d}(\frac{L}{\sqrt{N}}u_{i},\frac{L}{\sqrt{N}}\bar{u}_{i})=\frac{L^{2}}{N}%
H_{eff}(u_{i},\bar{u}_{i})=-\frac{1}{2}\frac{L^{2}}{N}\log \left\langle \psi
_{b}(u_{i},\bar{u}_{i})\mid \psi _{b}(u_{i},\bar{u}_{i})\right\rangle
\end{equation}%
where $\frac{L}{\sqrt{N}}u_{i}=z_{i},$ $K_{d}=K,~$and the $\psi _{b}(u_{i},%
\bar{u}_{i})$ denotes the reduced wavefunction for the test eigenvalue, and $%
H_{eff}$ here is the effective Hamiltonian for the test eigenvalue. We have $%
i=1,2,3$ for 1/8 BPS case. This suggests an interpretation that the $K_{d}$
is the effective Hamiltonian of a test eigenvalue. However, it may not be
the only interpretation.

The radial direction $r=\sqrt{\left\vert z_{1}\right\vert ^{2}+\left\vert
z_{2}\right\vert ^{2}+\left\vert z_{3}\right\vert ^{2}}$ in the 6d space is
already the $AdS$ radial direction $\cosh \rho $. The $S^{5}$ directions can
already be described by $r=1$. However, in the 6d space, there is more than
the $r=1~$region. When we put an eigenvalue in the location say$~r>1$, it is
a configuration that also describes the radial direction of $AdS$. So the
radial direction of the 6d space already describes the radial direction of
the $AdS$. The 6d space may be considered as $R^{6}\setminus B^{6}$.

One can also add ripples on the $S^{5}.$ As shown in \cite{Chen:2007du}, the
gauged supergravity solution of \cite{Chong:2004ce} for the scalar-gauge
system in 5D corresponds to ellipsoidal droplets, i.e. ripples on the $S^{5}$
with lowest possible wave-number.\

1/2 BPS geometries were also embedded in the 1/8 BPS ansatz, as performed in
\cite{Chen:2007du}, demonstrating topologically more nontrivial 5-surfaces.%
\vspace{1pt}

\subsection{More general cases}

\vspace{1pt}\label{sec:1-8_JJJ_more_general}

The K\"{a}hler potential for $AdS$ is in the general form

\begin{equation}
K={\textstyle}\frac{1}{2}f(z_{1},\bar{z}_{1})e^{\Xi (z_{i},\bar{z}_{i})}-{%
\textstyle}\frac{1}{2}\Xi (z_{i},\bar{z}_{i})-{\textstyle}\frac{1}{2}\log
f(z_{1},\bar{z}_{1})  \label{K_JJJ_ge_01}
\end{equation}%
$i=2,3.$ The $K$ of $AdS$ is isotropic in the 6d space. We now relax the
symmetry for $K$ on the 6d base space. We can first consider that the $z_{1},%
\bar{z}_{1}$ space and $z_{i},\bar{z}_{i}$ space ($i=2,3$ here) are not on
the equal footing.

The force balance equations are, e.g.
\begin{eqnarray}
&&-\partial _{1}K=\frac{1}{2}\partial _{1}f(e^{\Xi (z_{i},\bar{z}%
_{i})}-f^{-1})=0, \\
&&-\partial _{i}K=\frac{1}{2}\partial _{i}\Xi (fe^{\Xi (z_{i},\bar{z}%
_{i})}-1)=0,
\end{eqnarray}%
where $i=2,3.~$The above can be simultaneously solved by%
\begin{equation}
f(z_{1},\bar{z}_{1})=e^{-\Xi (z_{i},\bar{z}_{i})}.
\label{force_balance_5-surface}
\end{equation}%
In other words, we may consider (\ref{force_balance_5-surface}) as the
constraint giving the shape of the 5-surface argued in \cite{Chen:2007du}.

As a special case, one may consider
\begin{equation}
f(z_{1},\bar{z}_{1})=|z_{1}|^{2}
\end{equation}%
which in general means that the Kahler potential preserves $S^{1}~$symmetry
in the 6d base, and the surface would correspond to
\begin{equation}
|z_{1}|^{2}=e^{-\Xi (z_{i},\bar{z}_{i})}.
\end{equation}

\vspace{1pt}We may also consider a more general form
\begin{equation}
K=\frac{1}{2}a^{2}(z_{i},\bar{z}_{i})-\frac{1}{2}\log a^{2}(z_{i},\bar{z}%
_{i}),~~i=1,2,3  \label{K_JJJ_ge_02}
\end{equation}%
i.e. the $K$ is anisotropic in the 6d space. For example,%
\begin{equation}
a^{2}(z_{i},\bar{z}_{i})=(\left\vert z_{1}\right\vert ^{2}+\left\vert
z_{2}\right\vert ^{2}+\left\vert z_{2}\right\vert ^{2})\mathrm{exp}(\tilde{f}%
\sum_{l_{1},l_{2},l_{3}\in Z^{+}}\epsilon
_{l_{1},l_{2},l_{3}}(z_{1}^{l_{1}}z_{2}^{l_{2}}z_{3}^{l_{3}}+c.c.)).
\label{1-8_ripple}
\end{equation}

The force balance equations are
\begin{equation}
-\partial _{i}K=\frac{1}{2}(\partial _{i}\log
a^{2})(a^{2}-1)=0,~~~~-\partial _{\bar{i}}K=\frac{1}{2}(\partial _{\bar{i}%
}\log a^{2})(a^{2}-1)=0
\end{equation}%
which can be simultaneously solved by
\begin{equation}
a^{2}(z_{i},\bar{z}_{i})=1.
\end{equation}

From this, we see that the configuration (\ref{1-8_ripple}) corresponds to
adding ripples
\begin{equation}
\left\vert z_{1}\right\vert ^{2}+\left\vert z_{2}\right\vert ^{2}+\left\vert
z_{2}\right\vert ^{2}=1-\tilde{f}\sum_{l_{1},l_{2},l_{3}\in Z^{+}}\epsilon
_{l_{1},l_{2},l_{3}}(z_{1}^{l_{1}}z_{2}^{l_{2}}z_{3}^{l_{3}}+c.c.)
\end{equation}%
where the $\epsilon $'s are small, which may correspond to the operators
schematically
\begin{equation}
O\sim {{\prod_{l_{1},l_{2},l_{3}}e}}^{\mathrm{tr}%
(Z{}^{l_{1}}Y{}^{l_{2}}X^{l_{3}})}.
\end{equation}

More generally, if we start with
\begin{equation}
a^{2}(z_{i},\bar{z}_{i})=\widetilde{a}^{2}(z_{i},\bar{z}_{i})\mathrm{exp}(%
\tilde{f}\sum_{l_{1},l_{2},l_{3}\in Z^{+}}\epsilon
_{l_{1},l_{2},l_{3}}(z_{1}^{l_{1}}z_{2}^{l_{2}}z_{3}^{l_{3}}+c.c.)),
\end{equation}%
the constraint is then
\begin{equation}
\widetilde{a}^{2}(z_{i},\bar{z}_{i})=1-\tilde{f}\sum_{l_{1},l_{2},l_{3}\in
Z^{+}}\epsilon
_{l_{1},l_{2},l_{3}}(z_{1}^{l_{1}}z_{2}^{l_{2}}z_{3}^{l_{3}}+c.c.).
\end{equation}%
This would correspond to adding small ripples on a more general droplet, and
may correspond to schematically
\begin{equation}
O\sim {{\prod_{l_{1},l_{2},l_{3}}e}}^{\text{tr}%
(Z{}^{l_{1}}Y{}^{l_{2}}X^{l_{3}})}O_{\chi }
\end{equation}%
in which case $O_{\chi }$, or the superposition thereof, may correspond to
the droplet without adding the extra ripples.\vspace{1pt}

\section{1/4 BPS geometries with $S_{1},J$}

\vspace{1pt}\label{sec:1-4_SJ} 

\vspace{1pt}

\subsection{General ansatz}

As we discussed at the beginning of section \ref{sec:1-4_JJ}, the ansatz (%
\ref{1-4_ansatz}) also describes another set of 1/4 BPS states,
corresponding to having an $AdS$ spin or an $SO(4)$ spin $S_{1}$.~This case
corresponds to that the $S^{3}$ in the ansatz (\ref{1-4_ansatz}) is to be in
the $S^{5}$ directions.

For this type of 1/4 BPS sector, the dual operators we will consider here
are of the schematic form

\begin{equation}
O\sim {{\prod_{i=1}^{m}}}\text{tr}(D_{+\dot{+}}^{n_{2i}}{}Z{}^{n_{1i}})
\end{equation}%
where $Z$ is a complex scalar of $\mathcal{N}$=4 SYM, and $D_{+\dot{+}}$ is
a derivative operator on $S^{3}$. The subscripts $+,\dot{+}~$denotes SU(2)$%
_{L}$ and SU(2)$_{R}~$indices. The operator $D_{+\dot{+}}$ carries SU(2)$%
_{L} $ and SU(2)$_{R}~$spins $\frac{1}{2},\frac{1}{2},$ and thus carries $%
S_{1}=S_{L}+S_{R}=1$,$~S_{2}=S_{L}-S_{R}=0.~$This is a class of 1/4 BPS
operators. The BPS bound is satisfied as
\begin{equation}
\Delta -S_{1}-J=0.
\end{equation}

In this case, the $Z$ field (related to $z_{1}~$space) and the derivative $%
D_{+\dot{+}}$ (related to $z_{2}~$space) are not on an equal footing, so one
does not expect a totally radial symmetry for the $K$ in the 4d base.

One can look at special cases that the K\"{a}hler potential is radially
symmetric in the $z_{1}~$space and $z_{2}~$space separately, e.g. one can
consider special forms
\begin{equation}
K=K(\left\vert z_{1}\right\vert ,\left\vert z_{2}\right\vert ,y).
\end{equation}%
In this case, the equation (\ref{MA__ne1}) is%
\begin{equation}
\partial _{\left\vert z_{1}\right\vert ^{2}}(\left\vert z_{1}\right\vert
^{2}\partial _{\left\vert z_{1}\right\vert ^{2}}K)\partial
_{|z_{2}|^{2}}(|z_{2}|^{2}\partial
_{|z_{2}|^{2}}K)-|z_{1}|^{2}|z_{2}|^{2}(\partial _{|z_{1}|^{2}}\partial
_{|z_{2}|^{2}}K)^{2}=(1-4y^{2}\partial _{y^{2}}^{2}K)\frac{y}{8\sqrt{e}}%
e^{2\partial _{y^{2}}K}  \label{eqn_two_radial}
\end{equation}%
where we also used $e^{D}=\frac{1}{4\sqrt{e}}$.

\vspace{1pt}If using the gauged ansatz with $n\eta =2~$in subsection \ref%
{sec:1-4_JJ_ansatz}, the 1/2 BPS geometries, including $AdS_{5}\times S^{5}$,%
$~$were already embedded in this sector as in section 6.3 and equation
(6.73) of \cite{Chen:2007du}, when choosing different $Z$ that exchanges the
role of two $S^{3}$s.

\subsection{AdS, 1st case}

\vspace{1pt}\label{sec:1-4_SJ_AdS_1st}

The embedding of these states in the ungauged ansatz with $n\eta =1~$were
analyzed in \cite{Lunin:2008tf}, and this subsection analyzes these cases in
more details.

As mentioned before, the embedding of both $AdS_{5}\times S^{5}$ and 1/2 BPS
geometries in the ansatz describing states with $S_{1},J$ were alternatively
obtained in the section 6.3 of \cite{Chen:2007du} via the gauged ansatz with
$n\eta =2$.

We first study the $AdS_{5}\times S^{5}$ case with ungauged ansatz $n\eta =1$%
. We begin with rewriting the metric on $AdS_{5}\times S^{5},$%
\begin{equation}
ds^{2}=-h^{-2}(dt+\omega )^{2}+h^{2}((Z+\frac{1}{2})^{{-1}}2\partial
_{i}\partial _{{\bar{j}}}Kdz^{i}d{\bar{z}}^{\bar{j}}+dy^{2})+y(e^{G}d\Omega
_{3}^{2}+e{^{-G}}(d\psi )^{2})  \label{metric_SJ_01}
\end{equation}%
\begin{eqnarray}
ds^{2} &=&-\cosh ^{2}\rho dt^{2}+d\rho ^{2}+\sinh ^{2}\rho \left[ \cos
^{2}\alpha d\widetilde{\phi }_{2}^{2}+d\alpha ^{2}\right] +d\theta ^{2}+\cos
^{2}\theta d\widetilde{\phi }_{1}^{2}  \notag \\
&&+(\sin ^{2}\theta d\Omega _{3}^{2}+\sinh ^{2}\rho \sin ^{2}\alpha d\psi
^{2}).  \label{metric_SJ_02}
\end{eqnarray}

By comparing the $d\Omega _{3}^{2}$ factors and $d\psi ^{2}$ factors, we get
\begin{eqnarray}
&&y=\sin \theta \sinh \rho \sin \alpha ,\quad ~~~~~~~~~~~~~~~~e^{G}=\frac{%
\sin \theta }{\sinh \rho \sin \alpha },\quad  \label{metric_components_01} \\
&&Z=-\frac{1}{2}\frac{\sinh ^{2}\rho \sin ^{2}\alpha -\sin ^{2}\theta }{%
(\sinh ^{2}\rho \sin ^{2}\alpha +\sin ^{2}\theta )},~~~h^{-2}=\sinh ^{2}\rho
\sin ^{2}\alpha +\sin ^{2}\theta .  \label{metric_components_02}
\end{eqnarray}

Similar to the 1/2 BPS case \cite{Lin:2004nb}, we make a shift of the
angular variables $\widetilde{\phi }_{1}=\phi _{1}+t,~\widetilde{\phi }%
_{2}=\phi _{2}-t$. This gives the mixing terms between time and angular
variables $\phi _{1},\phi _{2}~$on the base where $z_{1}=\left\vert
z_{1}\right\vert e^{i\phi _{1}},z_{2}=\left\vert z_{2}\right\vert e^{i\phi
_{2}}$. So the one form $\omega $ is
\begin{equation}
\omega =h^{2}(-\cos ^{2}\theta d\phi _{1}+\sinh ^{2}\rho \cos ^{2}\alpha
d\phi _{2})  \label{AltAdSOm-2}
\end{equation}%
and
\begin{equation}
dt+\omega =dt+h^{2}(-\cos ^{2}\theta d\phi _{1}+\sinh ^{2}\rho \cos
^{2}\alpha d\phi _{2}).  \label{dt_w_1st}
\end{equation}

The metric in the $y$ direction and 4d base is

\begin{eqnarray}
&&h^{2}((Z+\frac{1}{2})^{{-1}}2\partial _{i}\partial _{{\bar{j}}}Kdz^{i}d{%
\bar{z}}^{\bar{j}}+dy^{2})=h^{2}dy^{2}  \notag \\
&&+\frac{2}{\sin ^{2}\theta }(\partial _{1}\partial _{\bar{1}}K(d\left\vert
z_{1}\right\vert ^{2}+\left\vert z_{1}\right\vert ^{2}d\phi
_{1}^{2})+\partial _{2}\partial _{\bar{2}}K(d\left\vert z_{2}\right\vert
^{2}+\left\vert z_{2}\right\vert ^{2}d\phi _{2}^{2})+\partial _{1}\partial _{%
\bar{2}}Kdz_{1}d{\bar{z}}_{2}+\partial _{2}\partial _{\bar{1}}Kdz_{2}d{\bar{z%
}}_{1})  \notag \\
&&
\end{eqnarray}%
\begin{eqnarray}
&=&d\rho ^{2}+d\theta ^{2}+\sinh ^{2}\rho d\alpha ^{2}+\frac{\cos ^{2}\theta
(\sinh ^{2}\rho \sin ^{2}\alpha +1)}{(\sinh ^{2}\rho \sin ^{2}\alpha +\sin
^{2}\theta )}d\phi _{1}^{2}+\frac{\sinh ^{2}\rho \cos ^{2}\alpha (\sinh
^{2}\rho +\sin ^{2}\theta )}{(\sinh ^{2}\rho \sin ^{2}\alpha +\sin
^{2}\theta )}d\phi _{2}^{2}  \notag \\
&&+\frac{2\sinh ^{2}\rho \cos ^{2}\alpha \cos ^{2}\theta }{(\sinh ^{2}\rho
\sin ^{2}\alpha +\sin ^{2}\theta )}d\phi _{1}d\phi _{2}.
\end{eqnarray}%
The $\rho ,\theta ,\alpha ~$directions are orthogonal to $\phi _{1},\phi
_{2} $. The $\theta ,\alpha ~$angles are orthogonal to each other. We may
therefore assume the change of variables from $\rho ,\theta ,\alpha $ to $%
\left\vert z_{1}\right\vert ,\left\vert z_{2}\right\vert $:
\begin{equation}
\left\vert z_{1}\right\vert =\left\vert z_{1}\right\vert (\rho ,\theta
),\quad \left\vert z_{2}\right\vert =\left\vert z_{2}\right\vert (\rho
,\alpha ).
\end{equation}

By comparing $g_{\phi _{1}\phi _{1}}$ we get
\begin{equation}
\partial _{1}\partial _{\bar{1}}K=\frac{\cos ^{2}\theta \sin ^{2}\theta }{%
2\left\vert z_{1}\right\vert ^{2}}\frac{\sinh ^{2}\rho \sin ^{2}\alpha +1}{%
\sinh ^{2}\rho \sin ^{2}\alpha +\sin ^{2}\theta }.  \label{K11_01}
\end{equation}

While by comparing $g_{\phi _{2}\phi _{2}}$ we have%
\begin{equation}
\partial _{2}\partial _{\bar{2}}K=\sinh ^{2}\rho \frac{\cos ^{2}\alpha
(\sinh ^{2}\rho +\sin ^{2}\theta )}{(\sinh ^{2}\rho \sin ^{2}\alpha +\sin
^{2}\theta )}\frac{\sin ^{2}\theta }{2\left\vert z_{2}\right\vert ^{2}}.
\label{K22_01}
\end{equation}

By comparing $g_{\theta \theta }$ we get%
\begin{equation}
\partial _{1}\partial _{\bar{1}}K=\frac{\sin ^{4}\theta }{2}\frac{\sinh
^{2}\rho \sin ^{2}\alpha +1}{\sinh ^{2}\rho \sin ^{2}\alpha +\sin ^{2}\theta
}(\frac{\partial \left\vert z_{1}\right\vert }{\partial \theta })^{-2}.
\label{K11_02}
\end{equation}%
Comparing (\ref{K11_01}),(\ref{K11_02}), we see that $(\log \left\vert
z_{1}\right\vert )^{2}=\log ^{2}(\cosh \rho \cos \theta ).~$It implies that
we have two possibilities
\begin{equation}
\left\vert z_{1}\right\vert =\cosh \rho \cos \theta  \label{z1_1st}
\end{equation}
or
\begin{equation}
\left\vert z_{1}\right\vert =\frac{1}{\cosh \rho \cos \theta }
\label{z1_2nd}
\end{equation}%
which are related by an inversion $\left\vert z_{1}\right\vert \rightarrow
\frac{1}{\left\vert z_{1}\right\vert }.$

We will only consider the first case (\ref{z1_1st}) in this subsection. In
this case (\ref{z1_1st}), when $\sin \theta =0$, the $S^{3}$ shrinks in $%
\left\vert z_{1}\right\vert =\cosh \rho \geqslant 1,$ which is the outside
part of a disk in $z_{1}$ space. When $\sinh \rho =0$, the $S^{1}$ shrinks
in $\left\vert z_{1}\right\vert =\cos \theta \leqslant 1$, which is the
inside part of a disk in $z_{1}$ space. The second case (\ref{z1_2nd}) will
be considered in subsection \ref{sec:1-4_SJ_AdS_2nd}.

By comparing $g_{\alpha \alpha }$ we have%
\begin{equation}
\partial _{2}\partial _{\bar{2}}K=\frac{\sin ^{2}\theta }{2}\frac{\sinh
^{2}\rho \sin ^{2}\alpha (\sinh ^{2}\rho +\sin ^{2}\theta )}{\sinh ^{2}\rho
\sin ^{2}\alpha +\sin ^{2}\theta }(\frac{\partial \left\vert
z_{2}\right\vert }{\partial \alpha })^{-2}.  \label{K22_02}
\end{equation}%
Comparing (\ref{K22_01}),(\ref{K22_02}), we see $(\log \left\vert
z_{2}\right\vert )^{2}=\log ^{2}(\tanh \rho \cos \alpha ).~$It also implies
that we have two possibilities, that are related by an inversion of the
radial variable. Without loss of generality, we consider the first case%
\begin{equation}
\left\vert z_{2}\right\vert =\tanh \rho \cos \alpha .
\end{equation}

Now we look at the AdS variables in terms of variables in the ansatz. From (%
\ref{metric_components_02}) we have
\begin{eqnarray}
&&Z=-\frac{1}{2}\frac{a^{2}+y^{2}-1}{\sqrt{(a^{2}+y^{2}-1)^{2}+4y^{2}}},
\label{Z_SJ} \\
&&a^{2}=\cos ^{2}\theta (1+\sinh ^{2}\rho \cos ^{2}\alpha )=\left\vert
z_{1}\right\vert ^{2}(1-\left\vert z_{2}\right\vert ^{2}).\quad
\label{a_1st}
\end{eqnarray}%
At $y=0$, $Z=-\frac{1}{2}~$where $a>1$, and $Z=\frac{1}{2}~$where $a<1$. So
this situation is opposite to the case (\ref{Z_JJ}) in section \ref%
{sec:1-4_JJ}. We can obtain $\rho ,\theta ,\alpha $ in terms of $%
z_{1},z_{2},y$ as e.g.
\begin{eqnarray}
\sinh ^{2}\rho \sin ^{2}\alpha &=&\frac{1}{2}(\sqrt{%
(a^{2}+y^{2}-1)^{2}+4y^{2}}+(a^{2}+y^{2}-1)), \\
\sin ^{2}\theta &=&\frac{1}{2}(\sqrt{(a^{2}+y^{2}-1)^{2}+4y^{2}}%
-(a^{2}+y^{2}-1)), \\
\cosh ^{2}\rho &=&\frac{1\text{+$\sinh ^{2}\rho \sin ^{2}\alpha $}}{%
(1-\left\vert z_{2}\right\vert ^{2})}.
\end{eqnarray}

Since
\begin{equation}
Z=-\frac{1}{2}y\partial _{y}(\frac{1}{y}\partial _{y}K)  \label{Z_K}
\end{equation}%
we can integrate $Z$ and write%
\begin{equation}
K=K_{Z}+\widetilde{K}_{0}+y^{2}\widetilde{K}_{1}  \label{K_sum}
\end{equation}%
where $K_{Z}$ is the result of integration of $Z$ in (\ref{Z_K})%
\begin{eqnarray}
K_{Z} &=&\frac{1}{4}y^{2}\log y^{2}+\frac{1}{4}(-b^{2}+(y^{2}+2)\log
(a^{2}+b^{2}+y^{2}+1))  \notag \\
&&-\frac{1}{4}y^{2}\log [2((a^{2}+b^{2}-1)b^{2}-(a^{2}+y^{2}+1)y^{2})],
\label{K_Z} \\
b^{2} &=&\sqrt{(a^{2}+y^{2}-1)^{2}+4y^{2}},~~~\ a^{2}=\left\vert
z_{1}\right\vert ^{2}(1-\left\vert z_{2}\right\vert ^{2}),
\end{eqnarray}%
and $\widetilde{K}_{0},\widetilde{K}_{1}$ are two integration constants that
do not depend on $y.$

By integrating the equations from the metric (\ref{K11_01}),(\ref{K22_01})
we get

\begin{eqnarray}
&&\partial _{\left\vert z_{1}\right\vert }K=\frac{-1+y^{2}+\left\vert
z_{1}\right\vert ^{2}(1-\left\vert z_{2}\right\vert ^{2})-\sqrt{%
(-1+y^{2}+\left\vert z_{1}\right\vert ^{2}(1-\left\vert z_{2}\right\vert
^{2}))^{2}+4y^{2}}}{4\left\vert z_{1}\right\vert ^{2}}, \\
&&\partial _{\left\vert z_{2}\right\vert }K=\frac{1+y^{2}-\left\vert
z_{1}\right\vert ^{2}(1-\left\vert z_{2}\right\vert ^{2})+\sqrt{%
(-1+y^{2}+\left\vert z_{1}\right\vert ^{2}(1-\left\vert z_{2}\right\vert
^{2}))^{2}+4y^{2}}}{4(1-\left\vert z_{2}\right\vert ^{2})}.
\end{eqnarray}%
Integrating these equations and comparing them with (\ref{K_sum}),(\ref{K_Z}%
), we obtain $\allowbreak $%
\begin{eqnarray}
&&\widetilde{K}_{0}+y^{2}\widetilde{K}_{1}=\frac{1}{4}(a^{2}-2\log a^{2})+%
\frac{1}{4}y^{2}(\log \left\vert z_{1}\right\vert ^{2}+\log 2e-\log
(1-\left\vert z_{2}\right\vert ^{2})),  \notag \\
&&a^{2}=\left\vert z_{1}\right\vert ^{2}(1-\left\vert z_{2}\right\vert ^{2}).
\end{eqnarray}%
So we get the final expression of $K,$
\begin{eqnarray}
K &=&\frac{1}{4}(a^{2}-2\log a^{2})+\frac{1}{4}y^{2}(\log \left\vert
z_{1}\right\vert ^{2}+\log 2e-\log (1-\left\vert z_{2}\right\vert ^{2}))
\notag \\
&&+\frac{1}{4}y^{2}\log y^{2}+\frac{1}{4}(-b^{2}+(y^{2}+2)\log
(a^{2}+b^{2}+y^{2}+1))  \notag \\
&&-\frac{1}{4}y^{2}\log [2((a^{2}+b^{2}-1)b^{2}-(a^{2}+y^{2}+1)y^{2})],
\label{K_1st_final} \\
b^{2} &=&\sqrt{(a^{2}+y^{2}-1)^{2}+4y^{2}},~~~\ a^{2}=\left\vert
z_{1}\right\vert ^{2}(1-\left\vert z_{2}\right\vert ^{2}).
\end{eqnarray}%
Now we check the one-form equation%
\begin{eqnarray}
\omega &=&\frac{1}{2y}\partial _{y}\left[ \left\vert z_{1}\right\vert
\partial _{\left\vert z_{1}\right\vert }Kd\phi _{1}+\left\vert
z_{2}\right\vert \partial _{\left\vert z_{2}\right\vert }Kd\phi _{2}\right]
\label{AltAdSOm-1-1-2} \\
&=&h^{2}(-\cos ^{2}\theta d\phi _{1}+\sinh ^{2}\rho \cos ^{2}\alpha d\phi
_{2}).  \label{eq:-10}
\end{eqnarray}%
We have
\begin{equation}
\partial _{y^{2}}(\left\vert z_{1}\right\vert \partial _{\left\vert
z_{1}\right\vert }K)=\frac{1}{2}(1-\frac{y^{2}+\left\vert z_{1}\right\vert
^{2}(1-\left\vert z_{2}\right\vert ^{2})+1}{\sqrt{(y^{2}+\left\vert
z_{1}\right\vert ^{2}(1-\left\vert z_{2}\right\vert ^{2})-1)^{2}+4y^{2}}}).
\label{AltAdSOm-1-1-1-1-1-1-1-1-1-1-1-1-1}
\end{equation}%
This exactly agrees with $-h^{2}\cos ^{2}\theta =-(Z+\frac{1}{2})\frac{\cos
^{2}\theta }{\sin ^{2}\theta }$.

We have
\begin{equation}
\partial _{y^{2}}(\left\vert z_{2}\right\vert \partial _{\left\vert
z_{2}\right\vert }K)=\frac{\left\vert z_{2}\right\vert ^{2}}{2(1-\left\vert
z_{2}\right\vert ^{2})}(1+\frac{y^{2}+\left\vert z_{1}\right\vert
^{2}(1-\left\vert z_{2}\right\vert ^{2})+1}{\sqrt{(y^{2}+\left\vert
z_{1}\right\vert ^{2}(1-\left\vert z_{2}\right\vert ^{2})-1)^{2}+4y^{2}}}).
\label{AltAdSOm-1-1-1-1-1-1-1-1-1-1-1-1-1-1}
\end{equation}%
This also exactly agrees with $h^{2}\sinh ^{2}\rho \cos ^{2}\alpha $.

We also checked that the final expression (\ref{K_1st_final}) exactly
satisfies (\ref{MA__ne1}).

For $AdS$ ground state, the droplet boundary is described by
\begin{equation}
a^{2}=\left\vert z_{1}\right\vert ^{2}(1-\left\vert z_{2}\right\vert
^{2})=1,\qquad \left\vert z_{2}\right\vert \leqslant 1.
\label{droplet_boundary_1st}
\end{equation}%
Since
\begin{equation}
\left\vert z_{2}\right\vert \leqslant 1
\end{equation}%
the $z_{2}$ space is a disk $D^{2}.$ The $z_{1}$ space has no restriction
and is $R^{2}.~$Thereby the 4d droplet space is $D^{2}\times R^{2}.~$The
3-surface in $D^{2}\times R^{2}~$described by (\ref{droplet_boundary_1st})
has $S_{\phi _{2}}^{1}\times S_{\phi _{1}}^{1}~$symmetry, where the $S_{\phi
_{2}}^{1}$ corresponds to the angle in the $S^{3}~$directions of $AdS$.
Thereby for more general solutions relaxing this symmetry, one can reduce
the spherical symmetry in the $AdS$ directions.

The $S^{1}\times S^{1}$ symmetry of the the ground state in this case, is
smaller than the $S^{3}$ symmetry of the ground state in the case in section %
\ref{sec:1-4_JJ}. \ \

\subsection{AdS, 2nd case}

\vspace{1pt}\label{sec:1-4_SJ_AdS_2nd}

In this subsection, we describe another embedding of the $AdS_{5}\times
S^{5} $, which is related to the one in subsection \ref{sec:1-4_SJ_AdS_1st}
via the inversion of $\left\vert z_{1}\right\vert $. The ansatz for the
metric components is written in the same way as in (\ref{metric_SJ_01}), (%
\ref{metric_SJ_02}), (\ref{metric_components_01}), (\ref%
{metric_components_02}) in subsection \ref{sec:1-4_SJ_AdS_1st}.

In this case, we make the angle shift $\widetilde{\phi }_{1}=\phi _{1}-t,~%
\tilde{t}=t+2\phi _{1},~\widetilde{\phi }_{2}=\phi _{2}-t$, so
\begin{eqnarray}
d\tilde{t}+\tilde{\omega} &=&dt+\omega ,\;\tilde{\omega}=\omega -2d\phi _{1}
\label{AltAdSOm-2-1} \\
\tilde{\omega} &=&h^{2}(\cos ^{2}\theta d\phi _{1}+\sinh ^{2}\rho \cos
^{2}\alpha d\phi _{2})-2d\phi _{1}  \label{tilde_w_2nd} \\
&=&\frac{1}{2y}\partial _{y}\left[ \left\vert z_{1}\right\vert \partial
_{\left\vert z_{1}\right\vert }Kd\phi _{1}+\left\vert z_{2}\right\vert
\partial _{\left\vert z_{2}\right\vert }Kd\phi _{2}\right] .
\label{tilde_w_2nd_02}
\end{eqnarray}%
We have
\begin{equation}
d\tilde{t}+\tilde{\omega}=dt+\omega =dt+h^{2}(\cos ^{2}\theta d\phi
_{1}+\sinh ^{2}\rho \cos ^{2}\alpha d\phi _{2}).  \label{dt_w_2nd}
\end{equation}%
So comparing the above equation (\ref{dt_w_2nd}), with (\ref{dt_w_1st}), the
net change is the reversal of the sign of $d\phi _{1}~$as in (\ref{dt_w_2nd}%
). The shift in $\tilde{\omega}$ is only an exact form. The symbolic
expressions for (\ref{K11_01}), (\ref{K22_01}), (\ref{K11_02}), (\ref{K22_02}%
) in terms of the $\rho ,\theta ,\alpha $ variables have no change. The
change is that we select the solution (\ref{z1_2nd}), i.e.%
\begin{equation}
\left\vert z_{1}\right\vert =\frac{1}{\cosh \rho \cos \theta }.
\label{z1_2nd_in_2nd_section}
\end{equation}%
In this case (\ref{z1_2nd_in_2nd_section}), when $\sin \theta =0$, the $%
S^{3} $ shrinks in $\left\vert z_{1}\right\vert =\frac{1}{\cosh \rho }%
\leqslant 1,$ which is the inside part of a disk in $z_{1}$ space. When $%
\sinh \rho =0$, the $S^{1}$ shrinks in $\left\vert z_{1}\right\vert =\frac{1%
}{\cos \theta }\geqslant 1$, which is the outside part of a disk in $z_{1}$
space. The inversion $\left\vert z_{1}\right\vert \rightarrow \frac{1}{%
\left\vert z_{1}\right\vert }$ exchanges the inside part with the outside
part of the disk in $z_{1}$ space.

\vspace{1pt}The expression of $a^{2}=\cos ^{2}\theta (1+\sinh ^{2}\rho \cos
^{2}\alpha )$ in terms of $\rho ,\theta ,\alpha $ variables, as in (\ref%
{a_1st}), has no change, and its dependence on $z_{1}~$changes to%
\begin{equation}
a^{2}=\cos ^{2}\theta (1+\sinh ^{2}\rho \cos ^{2}\alpha )=\frac{%
(1-\left\vert z_{2}\right\vert ^{2})}{\left\vert z_{1}\right\vert ^{2}}.
\label{a_2nd}
\end{equation}

After using the new expression (\ref{z1_2nd_in_2nd_section}) in (\ref{K11_01}%
),(\ref{K22_01}), and performing integrations of (\ref{K11_01}),(\ref{K22_01}%
), we get
\begin{eqnarray}
&&\partial _{\left\vert z_{1}\right\vert }K=\frac{-\frac{(1-\left\vert
z_{2}\right\vert ^{2})}{\left\vert z_{1}\right\vert ^{2}}+1-5y^{2}+\sqrt{%
(y^{2}+\frac{(1-\left\vert z_{2}\right\vert ^{2})}{\left\vert
z_{1}\right\vert ^{2}}-1)^{2}+4y^{2}}}{4\left\vert z_{1}\right\vert ^{2}},
\label{K_z1_2nd} \\
&&\partial _{\left\vert z_{2}\right\vert }K=\frac{-\frac{(1-\left\vert
z_{2}\right\vert ^{2})}{\left\vert z_{1}\right\vert ^{2}}+1+y^{2}+\sqrt{%
(y^{2}+\frac{(1-\left\vert z_{2}\right\vert ^{2})}{\left\vert
z_{1}\right\vert ^{2}}-1)^{2}+4y^{2}}}{4(1-\left\vert z_{2}\right\vert ^{2})}%
.  \label{K_z2_2nd}
\end{eqnarray}%
\begin{eqnarray}
K_{Z} &=&\frac{1}{4}y^{2}\log y^{2}+\frac{1}{4}(-b^{2}+(y^{2}+2)\log
(a^{2}+b^{2}+y^{2}+1))  \notag \\
&&-\frac{1}{4}y^{2}\log [2((a^{2}+b^{2}-1)b^{2}-(a^{2}+y^{2}+1)y^{2})],
\label{KZ_SJ_2nd} \\
b^{2} &=&\sqrt{(a^{2}+y^{2}-1)^{2}+4y^{2}},~~~\ a^{2}=\frac{(1-\left\vert
z_{2}\right\vert ^{2})}{\left\vert z_{1}\right\vert ^{2}}.
\end{eqnarray}%
After comparing (\ref{KZ_SJ_2nd}),(\ref{K_z1_2nd}),(\ref{K_z2_2nd}) with (%
\ref{K_sum}), we obtain%
\begin{equation}
\widetilde{K}_{0}+y^{2}\widetilde{K}_{1}=\frac{1}{4}(a^{2}-2\log a^{2})+%
\frac{1}{4}y^{2}(-\log \left\vert z_{1}\right\vert ^{2}+\log 2e-\log
(1-\left\vert z_{2}\right\vert ^{2}))-y^{2}\log \left\vert z_{1}\right\vert
^{2}.
\end{equation}%
So the final result for $K$ is
\begin{eqnarray}
K &=&\frac{1}{4}(a^{2}-2\log a^{2})+\frac{1}{4}y^{2}(-\log \left\vert
z_{1}\right\vert ^{2}+\log 2e-\log (1-\left\vert z_{2}\right\vert
^{2}))-y^{2}\log \left\vert z_{1}\right\vert ^{2}  \notag \\
&&+\frac{1}{4}y^{2}\log y^{2}+\frac{1}{4}(-b^{2}+(y^{2}+2)\log
(a^{2}+b^{2}+y^{2}+1))  \notag \\
&&-\frac{1}{4}y^{2}\log [2((a^{2}+b^{2}-1)b^{2}-(a^{2}+y^{2}+1)y^{2})],
\label{K_final_SJ_2nd} \\
b^{2} &=&\sqrt{(a^{2}+y^{2}-1)^{2}+4y^{2}},~~~\ a^{2}=\frac{(1-\left\vert
z_{2}\right\vert ^{2})}{\left\vert z_{1}\right\vert ^{2}}.
\end{eqnarray}

\vspace{1pt}We now check the one-form equation in (\ref{tilde_w_2nd}),(\ref%
{tilde_w_2nd_02}). We get from the final expression of $K$,%
\begin{eqnarray}
&&\partial _{y^{2}}(\left\vert z_{1}\right\vert \partial _{\left\vert
z_{1}\right\vert }K)=\frac{\frac{(1-\left\vert z_{2}\right\vert ^{2})}{%
\left\vert z_{1}\right\vert ^{2}}+1+y^{2}-\sqrt{(y^{2}+\frac{(1-\left\vert
z_{2}\right\vert ^{2})}{\left\vert z_{1}\right\vert ^{2}}-1)^{2}+4y^{2}}}{2%
\sqrt{(y^{2}+\frac{(1-\left\vert z_{2}\right\vert ^{2})}{\left\vert
z_{1}\right\vert ^{2}}-1)^{2}+4y^{2}}}-2,  \label{w_z1_2nd} \\
&&\partial _{y^{2}}(\left\vert z_{2}\right\vert \partial _{\left\vert
z_{2}\right\vert }K)=\frac{\left\vert z_{2}\right\vert ^{2}(\frac{%
(1-\left\vert z_{2}\right\vert ^{2})}{\left\vert z_{1}\right\vert ^{2}}%
+1+y^{2}+\sqrt{(y^{2}+\frac{(1-\left\vert z_{2}\right\vert ^{2})}{\left\vert
z_{1}\right\vert ^{2}}-1)^{2}+4y^{2}})}{2(1-\left\vert z_{2}\right\vert ^{2})%
\sqrt{(y^{2}+\frac{(1-\left\vert z_{2}\right\vert ^{2})}{\left\vert
z_{1}\right\vert ^{2}}-1)^{2}+4y^{2}}},
\end{eqnarray}%
and they agree with the expression of $\tilde{\omega}~$in (\ref{tilde_w_2nd}%
),(\ref{tilde_w_2nd_02}). We also checked that the final expression (\ref%
{K_final_SJ_2nd}) exactly satisfies (\ref{MA__ne1}).

For $AdS$ ground state, the droplet boundary is described by
\begin{equation}
a^{2}=(1-\left\vert z_{2}\right\vert ^{2})/\left\vert z_{1}\right\vert
^{2}=1~\quad \mathrm{i.e.}~\left\vert z_{2}\right\vert ^{2}+\left\vert
z_{1}\right\vert ^{2}=1;\qquad \left\vert z_{2}\right\vert \leqslant 1.
\label{surface_SJ_2nd}
\end{equation}%
Due to
\begin{equation}
\left\vert z_{2}\right\vert \leqslant 1
\end{equation}%
the droplet space is $D^{2}\times R^{2}.~$Although the 3-surface in (\ref%
{surface_SJ_2nd}) appears to be $S^{3}$ in $D^{2}\times R^{2}$, it does not
have an exact $S^{3}$ symmetry, since it only has the symmetry of $S_{\phi
_{2}}^{1}\times S_{\phi _{1}}^{1}.~$It is related to the first case in (\ref%
{droplet_boundary_1st}) in subsection \ref{sec:1-4_SJ_AdS_1st} by exchanging
the inside part with the outside part of a disk in the $R^{2}$ space of $%
z_{1},~$i.e.$~\left\vert z_{1}\right\vert \rightarrow \frac{1}{\left\vert
z_{1}\right\vert }.$

\subsection{Inversion and shift}

\vspace{1pt}\label{sec:1-4_SJ_inversion_shift}

In subsections \ref{sec:1-4_SJ_AdS_1st},\ref{sec:1-4_SJ_AdS_2nd}, we see
that there are two embeddings that are related by an inversion $%
|z_{1}|\rightarrow \frac{1}{|z_{1}|}$ , and also a shift of K\"{a}hler
potential at the same time. We denotes (\ref{K_1st_final}),(\ref%
{K_final_SJ_2nd}) as $K,\widetilde{K}~$respectively, in this subsection.
When comparing (\ref{K_1st_final}),(\ref{K_final_SJ_2nd}), we have
\begin{equation}
\widetilde{K}(\left\vert \widetilde{z}_{1}\right\vert ,\left\vert
z_{2}\right\vert ,y)=K(\left\vert z_{1}\right\vert =\frac{1}{\left\vert
\widetilde{z}_{1}\right\vert },\left\vert z_{2}\right\vert ,y)-y^{2}\log
\left\vert \widetilde{z}_{1}\right\vert ^{2}.  \label{relation_inversion}
\end{equation}

It's straightforward to see that this is correct from the equation (\ref%
{eqn_two_radial}). The (\ref{eqn_two_radial}) for $K$ is%
\begin{equation}
\frac{1}{\left\vert z_{1}\right\vert ^{2}\left\vert z_{2}\right\vert ^{2}}%
(\partial _{\log \left\vert z_{1}\right\vert ^{2}}^{2}K\partial _{\log
\left\vert z_{2}\right\vert ^{2}}^{2}K-(\partial _{\log \left\vert
z_{1}\right\vert ^{2}}\partial _{\log \left\vert z_{2}\right\vert
^{2}}K)^{2})=(1-4y^{2}\partial _{y^{2}}^{2}K)\frac{y}{8\sqrt{e}}e^{2\partial
_{y^{2}}K}.  \label{eqn_K_log}
\end{equation}%
Under the transform $\log \left\vert \widetilde{z}_{1}\right\vert =-\log
\left\vert z_{1}\right\vert ,$ and using (\ref{relation_inversion}) we have
\begin{eqnarray}
\frac{1}{\left\vert \widetilde{z}_{1}\right\vert ^{2}\left\vert
z_{2}\right\vert ^{2}}(\partial _{\log \left\vert \widetilde{z}%
_{1}\right\vert ^{2}}^{2}\widetilde{K}\partial _{\log \left\vert
z_{2}\right\vert ^{2}}^{2}\widetilde{K}-(\partial _{\log \left\vert
\widetilde{z}_{1}\right\vert ^{2}}\partial _{\log \left\vert
z_{2}\right\vert ^{2}}\widetilde{K})^{2}) &=&(1-4y^{2}\partial _{y^{2}}^{2}K)%
\frac{y}{8\sqrt{e}}e^{2\partial _{y^{2}}K}e^{2\partial
_{_{y^{2}}}(-y^{2}\log \left\vert \widetilde{z}_{1}\right\vert ^{2})}  \notag
\\
&=&(1-4y^{2}\partial _{y^{2}}^{2}\widetilde{K})\frac{y}{8\sqrt{e}}%
e^{2\partial _{y^{2}}\widetilde{K}},
\end{eqnarray}%
hence $\widetilde{K}$ also satisfy the equation.

The metric involves rescaling transformation, but it does not change the
gravity background essentially. However, the one-form $\tilde{\omega}$ has a
subtle shift by an exact form, as discussed in subsection \ref%
{sec:1-4_SJ_AdS_2nd}. This is due to the shift $-y^{2}\log \left\vert
\widetilde{z}_{1}\right\vert ^{2}~$in (\ref{relation_inversion}), and is
also the reason for the shift happened in (\ref{w_z1_2nd}).

There is also another shift symmetry: One can add to $K$ a term $-c_{1}\log
\left\vert z_{1}\right\vert ^{2}$ or $-c_{2}\log \left\vert z_{2}\right\vert
^{2}$, where $c_{1},c_{2}$ are constants, without changing the equation (\ref%
{eqn_K_log}), or changing the solution.

\subsection{Small $y$}

\vspace{1pt}In this subsection, we analyze the small $y$ behavior of general
geometries with $S_{1}~$and $J$.

The small $y$ expansion of the $AdS$ expression in subsection \ref%
{sec:1-4_SJ_AdS_1st} for $a^{2}=\left\vert z_{1}\right\vert
^{2}(1-\left\vert z_{2}\right\vert ^{2})~$is as follows:

When $S^{3}\rightarrow 0$, $Z=-\frac{1}{2},~$i.e. $a^{2}>1$,
\begin{eqnarray}
&&K=\frac{1}{4}y^{2}\mathrm{\log }%
y^{2}+K_{0}+y^{2}K_{1}+(y^{2})^{2}K_{2}+o((y^{2})^{3}),\; \\
&&K_{0}=\frac{1}{4}(1+2\log 2),~~~K_{1}=\frac{1}{2}(\log \left\vert
z_{1}\right\vert ^{2}-\log (a^{2}-1)),~~~K_{2}=-\frac{1}{4(a^{2}-1)^{2}}.
\end{eqnarray}

On the other hand, when $S^{1}\rightarrow 0$, $Z=\frac{1}{2},$ i.e. $a^{2}<1$%
,
\begin{eqnarray}
&&K=-\frac{1}{4}y^{2}\mathrm{\log }%
y^{2}+K_{0}+y^{2}K_{1}+(y^{2})^{2}K_{2}+o((y^{2})^{3}),\;
\label{K_S1_shrink_a} \\
&&K_{0}=\frac{1}{2}(a^{2}-\text{log}a^{2})+\frac{1}{4}(-1+2\log 2),~K_{1}=%
\frac{1}{2}(1-\log (1-\left\vert z_{2}\right\vert ^{2})+\log (1-a^{2})),
\notag \\
&&  \label{K_S1_shrink__a_02} \\
&&K_{2}=\frac{1}{4(a^{2}-1)^{2}},~~~\;e^{2K_{1}}=\frac{e(1-a^{2})}{%
(1-\left\vert z_{2}\right\vert ^{2})}.
\end{eqnarray}

While, very similarly, the small $y$ expansion of the $AdS$ expression in
subsection \ref{sec:1-4_SJ_AdS_2nd} for $a^{2}=\frac{(1-\left\vert
z_{2}\right\vert ^{2})}{\left\vert z_{1}\right\vert ^{2}}~$is:

When $S^{3}\rightarrow 0$, $Z=-\frac{1}{2},~$i.e. $a^{2}>1$,
\begin{eqnarray}
&&K=\frac{1}{4}y^{2}\mathrm{\log }%
y^{2}+K_{0}+y^{2}K_{1}+(y^{2})^{2}K_{2}+o((y^{2})^{3}),\; \\
&&K_{0}=\frac{1}{4}(1+2\log 2),~K_{1}=\frac{1}{2}(-\log \left\vert
z_{1}\right\vert ^{2}-\log (a^{2}-1))-\log \left\vert z_{1}\right\vert
^{2},~K_{2}=-\frac{1}{4(a^{2}-1)^{2}}.  \notag \\
&&
\end{eqnarray}

On the other hand, when $S^{1}\rightarrow 0$, $Z=\frac{1}{2},$ i.e. $a^{2}<1$%
,
\begin{eqnarray}
&&K=-\frac{1}{4}y^{2}\mathrm{\log }%
y^{2}+K_{0}+y^{2}K_{1}+(y^{2})^{2}K_{2}+o((y^{2})^{3}),\; \\
&&K_{0}=\frac{1}{2}(a^{2}-\text{log}a^{2})+\frac{1}{4}(-1+2\log 2),~K_{1}=%
\frac{1}{2}(1-\log (1-\left\vert z_{2}\right\vert ^{2})+\log (1-a^{2}))-\log
\left\vert z_{1}\right\vert ^{2},  \notag \\
&& \\
&&K_{2}=\frac{1}{4(a^{2}-1)^{2}},~~\ \ \ e^{2K_{1}}=\frac{e(1-a^{2})}{%
\left\vert z_{1}\right\vert ^{4}(1-\left\vert z_{2}\right\vert ^{2})}.\;
\end{eqnarray}

These solutions satisfies the small $y$ equations (\ref{small_y_minus_K2}),(%
\ref{small_y_minus_K0}),(\ref{small_y_minus_K1}),(\ref{small_y_plus_K0}),(%
\ref{small_y_plus_K1}) analyzed in subsection \ref{sec:1-4_JJ_small_y}, and
small $y$ regularity conditions (\ref{boundary_condition_plus}), (\ref%
{boundary_condition_minus}).

Now we study more general solutions in the first case of \ref%
{sec:1-4_SJ_AdS_1st}. We take both $\left\vert z_{2}\right\vert ^{2}\ll
1,\left\vert z_{1}\right\vert ^{2}\ll 1$.~Thus $a^{2}=\left\vert
z_{1}\right\vert ^{2}(1-\left\vert z_{2}\right\vert ^{2})\ll 1.$ This is
near the origin of the disk in the $z_{2}$ space, as well as near the origin
of the $Z=\frac{1}{2}~$droplet region. We have%
\begin{equation}
K_{0}=-{\textstyle\frac{1}{2}}\log \left\vert z_{1}\right\vert ^{2}+{%
\textstyle\frac{1}{2}}(\left\vert z_{1}\right\vert ^{2}+\left\vert
z_{2}\right\vert ^{2})+\widetilde{K}_{0},\qquad K_{1}={\textstyle\frac{1}{2}}%
+\widetilde{K}_{1}.
\end{equation}

As discussed in subsection \ref{sec:1-4_SJ_inversion_shift}, one can also
add $-\frac{1}{2}\log \left\vert z_{2}\right\vert ^{2}~$in$~K$, without
changing the solution. So one can also equivalently define
\begin{equation}
K_{0}={\textstyle\frac{1}{2}}\left\vert z_{1}\right\vert ^{2}-{\textstyle
\frac{1}{2}}\log \left\vert z_{1}\right\vert ^{2}+{\textstyle\frac{1}{2}}%
\left\vert z_{2}\right\vert ^{2}-{\textstyle\frac{1}{2}}\log \left\vert
z_{2}\right\vert ^{2}+\widetilde{K}_{0},\qquad K_{1}={\textstyle\frac{1}{2}}+%
\widetilde{K}_{1}.
\end{equation}

When $S^{1}\rightarrow 0$, $Z=\frac{1}{2},$ i.e. $a^{2}<1$, the small $y$
equations (\ref{small_y_plus_K0}),(\ref{small_y_plus_K1}) become, in the
leading order in tilded variables, as%
\begin{eqnarray}
&&\partial _{1}\partial _{\bar{1}}\widetilde{K}_{0}+\partial _{2}\partial _{%
\bar{2}}\widetilde{K}_{0}=\widetilde{K}_{1}, \\
&&\partial _{1}\partial _{\bar{1}}\widetilde{K}_{1}+\partial _{2}\partial _{%
\bar{2}}\widetilde{K}_{1}=0.
\end{eqnarray}

So we get in leading order in small $a~$region
\begin{eqnarray}
&&\widetilde{K}_{1}=-\frac{1}{2}\int_{{\normalcolor\mathcal{D}}}(\left\vert
z_{1}-z_{1}^{\prime }\right\vert ^{2}-\left\vert z_{2}-z_{2}^{\prime
}\right\vert ^{2})u(z_{i}^{\prime },\bar{z}_{i}^{\prime })\frac{1}{n}%
d^{2}z_{1}^{\prime }d^{2}z_{2}^{\prime }, \\
&&\widetilde{K}_{0}=-\frac{1}{2}\int_{{\normalcolor\mathcal{D}}}(\left\vert
z_{1}-z_{1}^{\prime }\right\vert ^{2}\left\vert z_{2}-z_{2}^{\prime
}\right\vert ^{2}-\frac{1}{2}\left\vert z_{2}-z_{2}^{\prime }\right\vert
^{4})u(z_{i}^{\prime },\bar{z}_{i}^{\prime })\frac{1}{n}d^{2}z_{1}^{\prime
}d^{2}z_{2}^{\prime }
\end{eqnarray}%
where $\int_{{\normalcolor\mathcal{D}}}u(z_{i}^{\prime },\bar{z}_{i}^{\prime
})\frac{1}{n}d^{2}z_{1}^{\prime }d^{2}z_{2}^{\prime }=1.$

\subsection{Large $y$}

In this subsection, we turn to the study of the large $y$ region of the
geometries with spin $S_{1}~$and $J$.

We first study the first case as in subsection \ref{sec:1-4_SJ_AdS_1st}. We
focus on large $y$ region and make a change of variable
\begin{equation}
K(z_{i},\bar{z}_{i},y)=\frac{1}{4}y^{2}\log y^{2}-\frac{1}{2}\log y^{2}-%
\frac{1}{2}y^{2}\log (1-\left\vert z_{2}\right\vert ^{2})-\frac{1}{2y^{2}}%
\left\vert z_{1}\right\vert ^{2}(1-\left\vert z_{2}\right\vert ^{2})+\frac{1%
}{2}\left\vert z_{2}\right\vert ^{2}+V(z_{i},\bar{z}_{i},y).
\end{equation}

After cancelling the leading terms in large $y$ in (\ref{MA__ne1}) exactly,
we get a linear equation%
\begin{equation}
\frac{1}{(1-\left\vert z_{2}\right\vert ^{2})}4\partial _{1}\partial _{\bar{1%
}}V+\frac{1}{y^{4}}(1-\left\vert z_{2}\right\vert ^{2})^{2}4\partial
_{2}\partial _{\bar{2}}V+y\partial _{y}(\frac{1}{y}\partial _{y}V)=0.
\end{equation}

If we further take a small $\left\vert z_{2}\right\vert \ll 1~$limit, we get%
\begin{equation}
4\partial _{1}\partial _{\bar{1}}V+y\partial _{y}(\frac{1}{y}\partial _{y}V)+%
\frac{1}{y^{4}}4\partial _{2}\partial _{\bar{2}}V=0.  \label{V_linear_SJ}
\end{equation}%
We can also change variables to%
\begin{eqnarray}
&&y^{2}\Psi =V, \\
&&4\partial _{1}\partial _{\bar{1}}\Psi +\frac{1}{y^{3}}\partial
_{y}(y^{3}\partial _{y}\Psi )+\frac{4}{y^{4}}\partial _{2}\partial _{\bar{2}%
}\Psi =0.  \label{Psi_linear_SJ}
\end{eqnarray}%
The first two terms are given by a 6d Laplacian; while near fixed $y$, the
equation is similar to an 8d Laplace equation.

The analysis for the second case as in subsection \ref{sec:1-4_SJ_AdS_2nd}
is very similar, and after the inversion \vspace{1pt}$|z_{1}|\rightarrow
\frac{1}{|z_{1}|},$ gives the same equations (\ref{V_linear_SJ}),(\ref%
{Psi_linear_SJ}), since the inversion does not affect the large $y$ or small
$\left\vert z_{2}\right\vert ~$limit.

If we expand (\ref{Psi_linear_SJ}) around a certain $y_{0}$, i.e. $%
y=y_{0}+x,~$we have
\begin{eqnarray}
&&\partial _{1}\partial _{\bar{1}}\Psi +\frac{1}{y_{0}^{4}}\partial
_{2}\partial _{\bar{2}}\Psi +\frac{1}{4}\partial _{x}^{2}\Psi =0. \\
&&\Psi =\int_{{\normalcolor\mathcal{D}}}\frac{u(z_{i}^{\prime },\bar{z}%
_{i}^{\prime })d^{2}z_{1}^{\prime }d^{2}z_{2}^{\prime }}{(\left\vert
z_{1}-z_{1}^{\prime }\right\vert ^{2}+\left\vert y-y_{0}\right\vert
^{2}+y_{0}^{4}\left\vert z_{2}-z_{2}^{\prime }\right\vert ^{2})^{\frac{3}{2}}%
}.
\end{eqnarray}%
This solution tells certain information that one can distribute droplets in
the $z_{i}^{\prime },\bar{z}_{i}^{\prime }$ space, where $i=1,2.$

\section{1/8 BPS geometries with $S_1, S_2, J$}

\label{sec:1-8_SSJ}

\subsection{General ansatz}

\vspace{1pt}\label{sec:1-8_SSJ_general_ansatz}

As we discussed at the beginning of section \ref{sec:1-8_JJJ}, the ansatz (%
\ref{1-8_ansatz}) also describes another set of 1/8 BPS states,
corresponding to having $AdS$ spins or $SO(4)$ spins $S_{1},S_{2}$.~This
case corresponds to that the $S^{3}$ in the ansatz (\ref{1-8_ansatz}) is to
be in the $S^{5}$ directions.

For this type of 1/8 BPS sector, since we have two $SO(4)$ spins $%
S_{1},S_{2} $,~we can consider a number of fields, such as the ones in table
(\ref{elements}),

\vspace{1pt}%
\begin{equation}
\begin{tabular}{lllll}
& $\Delta $ & $(S_{L},S_{R})$ & $(S_{1},S_{2})$ & $(H_{1},H_{2},H_{3})$ \\
\hline
$Z$ & 1 & (0,0) & (0,0) & (0,1,0) \\
$D_{+\dot{+}}$ & 1 & ($\frac{1}{2},\frac{1}{2}$) & (1,0) & (0,0,0) \\
$D_{+\dot{-}}$ & 1 & ($\frac{1}{2}$,-$\frac{1}{2}$) & (0,1) & (0,0,0) \\
$\lambda _{3+}$ & $\frac{3}{2}$ & ($\frac{1}{2},0$) & ($\frac{1}{2},\frac{1}{%
2}$) & (0,1,-1) \\
$\lambda _{4+}$ & $\frac{3}{2}$ & ($\frac{1}{2},0$) & ($\frac{1}{2},\frac{1}{%
2}$) & (0,0,1)%
\end{tabular}
\label{elements}
\end{equation}%
as well as the field strength \vspace{1pt}$F_{++}.$ We see that $D_{+\dot{+}%
}D_{+\dot{-}}Z$ has the same quantum numbers as$~\lambda _{3+}\lambda _{4+}$
and $ZF_{++}.$ One can consider representative operators of the schematic
form
\begin{equation}
O\sim {{\prod_{i=1}^{m}}}\mathrm{tr}(D_{+\dot{+}}^{n_{2i}}{}D_{+\dot{-}%
}^{n_{3i}}{}Z{}^{n_{1i}})
\end{equation}%
where some of the $D_{+\dot{+}}D_{+\dot{-}}Z$ may be replaced by $\lambda
_{3+}\lambda _{4+}$ or $ZF_{++}.$ The BPS bound is satisfied as

{\large
\begin{equation}
\Delta -S_{1}-S_{2}-J=0.
\end{equation}%
}

We have two spins in $AdS$ direction. In this case, the $Z$ field (related
to the $z_{1}~$space) and the derivative $D_{+\dot{+}},D_{+\dot{-}}$
(related to the $z_{2},z_{3}~$space) are not on an equal footing, so one
does not expect a totally radial symmetry for the $K$ in the 6d base.

\subsection{AdS, 1st case}

\vspace{1pt}\label{sec:1-8_SSJ_AdS_1st}

We first study the $AdS$ case. The embedding of the 1/2 BPS geometries in
the 1/8 BPS ansatz were obtained in section 5.4 of \cite{Chen:2007du}, with
the focus on the case that the $S^{3}~$of the 1/8 BPS ansatz is in the $AdS$
direction. The general formulas in \cite{Chen:2007du} are also applicable to
the case when the $S^{3}~$of the 1/8 BPS ansatz is in the $S^{5}$ direction.

The embedding is
\begin{eqnarray}
&&\log (r^{2}(z_{1},\bar{z}_{1},y))=\int^{y^{2}}\left( Z(z_{1},\bar{z}%
_{1},y^{\prime })+\frac{1}{2}\right) \frac{{d(y^{\prime }{}^{2})}}{{%
y^{\prime }{}^{2}}},  \label{eq:7rydef} \\
&&\frac{dr}{r}=-i(V_{z}dz_{1}-V_{\bar{z}}d\bar{z}_{1})+\frac{Z+\frac{1}{2}}{y%
}dy,
\end{eqnarray}%
\begin{equation}
K(z_{1},\bar{z}_{1},y^{2})=\frac{1}{2}\int^{y^{2}}(Z(z_{1},\bar{z}%
_{1},y^{\prime })+\frac{1}{2})d(y^{\prime }{}^{2}).  \label{eq:7kpotint}
\end{equation}%
where $r^{2}=|z_{2}|^{2}+|z_{3}|^{2}.~$The integrals will yield expressions
up to a term that only depends on $z_{1}$ directions, and is independent of $%
y$.

Now we embed $AdS,$ in the way that the $S^{3}~$in the ansatz is in the $%
S^{5}$. We consider the $Z$ as,
\begin{equation}
Z=-\frac{1}{2}\frac{|z_{1}|^{2}+y^{2}-1}{\sqrt{%
(|z_{1}|^{2}+y^{2}-1)^{2}+4y^{2}}}.
\end{equation}
Comparing to (5.104) of \cite{Chen:2007du}, the role of the two $S^{3}$'s
are switched for this alternative case. The variables are in the unit that
the $AdS$ radius $L$=1, and in the above the $y$ is the $y$ variable in the
ansatz of the 1/2 BPS geometries.

After performing the $y$ integral, we have
\begin{eqnarray}
K &=&\frac{1}{4}(y^{2}-\sqrt{(|z_{1}|^{2}+y^{2}-1)^{2}+4y^{2}})+\frac{1}{2}%
\log (1+|z_{1}|^{2}+y^{2}+\sqrt{(|z_{1}|^{2}+y^{2}-1)^{2}+4y^{2}}%
)+f_{1}(|z_{1}|),  \notag \\
&&  \label{K_AdS_SSJ}
\end{eqnarray}%
\begin{equation}
r^{4}=(\frac{2}{1-|z_{1}|^{2}}+\frac{4(|z_{1}|^{2}-y^{2}-1)}{%
(1-|z_{1}|^{2})^{2}(1+|z_{1}|^{2}+y^{2}+\sqrt{%
(|z_{1}|^{2}+y^{2}-1)^{2}+4y^{2}})})f_{2}(|z_{1}|)  \label{z2z3_ads_SSJ}
\end{equation}%
and we have
\begin{eqnarray}
&&f_{1}(|z_{1}|)=\frac{1}{4}(1+|z_{1}|^{2})-\frac{1}{2}\log |z_{1}|^{2}-%
\frac{1}{2}\log 2,  \label{AldAdSZ-1-1-1-1-3} \\
&&\log f_{2}(|z_{1}|)=-\frac{1}{4}\log (\frac{2|z_{1}|^{2}}{%
(1-|z_{1}|^{2})^{2}}).
\end{eqnarray}

Then we get%
\begin{eqnarray}
&&|z_{2}|^{2}=\tanh ^{2}\rho \cos ^{2}\alpha , \\
&&|z_{3}|^{2}=\tanh ^{2}\rho \sin ^{2}\alpha , \\
&&|z_{1}|^{2}=\cosh ^{2}\rho \cos ^{2}\theta .
\end{eqnarray}%
where the $\rho ,\theta ,\alpha ~$variables are defined in the same way as
in (\ref{metric_SJ_02}), and%
\begin{equation}
y_{1/8}^{2}=1-(1-|z_{2}|^{2}-|z_{3}|^{2})|z_{1}|^{2}.
\end{equation}%
where $y_{1/8}^{2}$ is the $y^{2}~$variable in the 1/8 BPS ansatz.

And the $y^{2}~$variable in the 1/2 BPS ansatz is
\begin{equation}
y_{1/2}^{2}=(\frac{1}{1-|z_{2}|^{2}-|z_{3}|^{2}}%
-|z_{1}|^{2})(|z_{2}|^{2}+|z_{3}|^{2}).
\end{equation}

Using these variables we simplify $K,$
\begin{equation}
K=\frac{1}{2}(1-|z_{2}|^{2}-|z_{3}|^{2})|z_{1}|^{2}-\frac{1}{2}\log
((1-|z_{2}|^{2}-|z_{3}|^{2})|z_{1}|^{2})  \label{K_AdS_SSJ_1st}
\end{equation}%
in the unit with $AdS$ radius $L$=1. The $K$ has an $S^{3}\times S^{1}~$%
symmetry.

In this case, we have
\begin{equation}
|z_{2}|^{2}+|z_{3}|^{2}\leqslant 1
\end{equation}%
Hence the $z_{2},z_{3}~$space is a $B^{4}$, whereas the $z_{1}~$space is $%
R^{2}.~$The droplet space thus would be $B^{4}\times R^{2}$.

The surface that $y_{1/8}^{2}=0$ is
\begin{equation}
|z_{2}|^{2}+|z_{3}|^{2}+|z_{1}|^{-2}=1
\end{equation}%
or%
\begin{equation}
|z_{1}|^{2}-|z_{1}|^{2}|z_{2}|^{2}-|z_{1}|^{2}|z_{3}|^{2}=1
\end{equation}%
If we focus near the origin of the $z_{2},z_{3}$ space, this will be near
the circle $|z_{1}|\approx 1~$in $z_{1}~$space$,$ and we see that $%
|z_{2}|^{2}+|z_{3}|^{2}\approx \epsilon ^{2}~$there (where $\epsilon ^{2}~$%
is small), so the $S^{3}~$in$~z_{2},z_{3}$ space shrinks at the boundary of $%
D^{2}$ in $z_{1}$ space. The surface$~$in $B^{4}\times R^{2}$ has $%
S^{3}\times S^{1}~$symmetry$.$

The equation
\begin{equation}
|z_{2}|^{2}+|z_{3}|^{2}=\tanh ^{2}\rho =\frac{|z_{1}|^{2}-1}{|z_{1}|^{2}}
\end{equation}%
suggests that we have an $S^{3}$ and its radius is not constant, and changes
with $|z_{1}|$. This means that the size of the $S^{3}~$changes with a
radial direction. The radius of $S^{3}$ and the radius of $S^{1},~$i.e. $%
|z_{1}|$ combine into the overall radial coordinate of the 6d space. In
other words, the radius of the $S^{3}$ can be viewed as the radial
coordinate of the 6d space, projected to the $(z_{2},z_{3})$ subspace.

If we recover the $AdS$ radius $L$ in the expressions, we have that (\ref%
{K_AdS_SSJ_1st}) gives
\begin{equation}
K=\frac{1}{2}\left\vert z_{1}\right\vert ^{2}(1-\frac{\left\vert
z_{2}\right\vert ^{2}}{L^{2}}-\frac{\left\vert z_{3}\right\vert ^{2}}{L^{2}}%
)-\frac{1}{2}L^{2}\log (1-\frac{\left\vert z_{2}\right\vert ^{2}}{L^{2}}-%
\frac{\left\vert z_{3}\right\vert ^{2}}{L^{2}})-\frac{1}{2}L^{2}\log
\left\vert z_{1}\right\vert ^{2}.  \label{K_AdS_SSJ_1st_L}
\end{equation}

Since the $S^{3}~$in the $z_{2},z_{3}$ space is the $S^{3}~$of $AdS$
directions, we can have excitations described by general configurations in $%
z_{2},z_{3}$ space that correspond to reducing the symmetry of $S^{3}~$in $%
AdS$ directions.

\subsection{AdS, 2nd case}

\vspace{1pt}\label{sec:1-8_SSJ_AdS_2nd}

In this subsection, we describe another embedding by inversion of $|z_{1}|$.
The initial steps are the same as in (\ref{eq:7rydef}),(\ref{eq:7kpotint}),(%
\ref{K_AdS_SSJ}),(\ref{z2z3_ads_SSJ}) in subsection \ref{sec:1-8_SSJ_AdS_1st}%
, with the difference of inverting $|z_{1}|$ as%
\begin{eqnarray}
&&|z_{2}|^{2}=\tanh ^{2}\rho \cos ^{2}\alpha ,\quad \\
&&|z_{3}|^{2}=\tanh ^{2}\rho \sin ^{2}\alpha , \\
&&|z_{1}|^{2}=\frac{1}{\cosh ^{2}\rho \cos ^{2}\theta }.
\label{z1_inversion_SSJ}
\end{eqnarray}%
\begin{equation}
y_{1/8}^{2}=1+\frac{(|z_{2}|^{2}+|z_{3}|^{2}-1)}{|z_{1}|^{2}}.
\end{equation}

Then we have in (\ref{K_AdS_SSJ}),(\ref{z2z3_ads_SSJ})
\begin{eqnarray}
&&f_{1}(|z_{1}|)=\frac{1}{4}(1+\frac{1}{|z_{1}|^{2}})+\frac{1}{2}\log
|z_{1}|^{2}-\frac{1}{2}\log 2,  \label{AldAdSZ-1-1-1-1-3-1} \\
&&\log f_{2}(|z_{1}|)=-\frac{1}{4}\log (\frac{2|z_{1}|^{2}}{%
(1-|z_{1}|^{2})^{2}}).
\end{eqnarray}

Using these variables we simplify $K$,
\begin{equation}
K=\frac{1}{2}\frac{(1-|z_{2}|^{2}-|z_{3}|^{2})}{|z_{1}|^{2}}-\frac{1}{2}\log
(1-|z_{2}|^{2}-|z_{3}|^{2})+\frac{1}{2}\log |z_{1}|^{2}.
\end{equation}%
in the unit with $AdS$ radius $L$=1. It has symmetry $S^{3}\times S^{1}$.

In this case, we also have
\begin{equation}
|z_{2}|^{2}+|z_{3}|^{2}\,\leqslant 1.
\end{equation}%
Hence the $z_{2},z_{3}~$space is a $B^{4}$, while the $z_{1}~$space is $%
R^{2}.~$The droplet space is also $B^{4}\times R^{2}$.

The surface that $y_{1/8}^{2}=0$ is
\begin{equation}
|z_{2}|^{2}+|z_{3}|^{2}+|z_{1}|^{2}=1.  \label{sp_SSJ_2nd}
\end{equation}%
If change of variable ($z_{2}/z_{1}\rightarrow z_{2},$$z_{3}/z_{1}%
\rightarrow z_{3}$),%
\begin{equation}
|z_{1}|^{2}+|z_{1}|^{2}|z_{2}|^{2}+|z_{1}|^{2}|z_{3}|^{2}=1.
\end{equation}%
Near the circle $|z_{1}|\approx 1,$ we see that $|z_{2}|^{2}+|z_{3}|^{2}%
\approx \epsilon ^{2},~$so the $S^{3}~$in$~z_{2},z_{3}$ space shrinks at the
boundary of $D^{2}$ in $z_{1}$ space. The surface in $B^{4}\times R^{2}~$may
be considered as the fibration of $S^{3}~$over a $D^{2}.~$The $K$ has an $%
S^{3}\times S^{1}~$symmetry. Although the equation resembles an $S^{5}$, it
may not describe a round sphere. Since the $z_{2},$$z_{3}~$space are not on
an equal footing with the $z_{1}~$space, and there is no symmetry between
the $z_{2},$$z_{3}~$space and the $z_{1}$$~$space, one may introduce
rescalings and (\ref{sp_SSJ_2nd}) would appear to describe only deformed $%
S^{5}~$in$~B^{4}\times R^{2}.~$

The equation
\begin{equation}
|z_{2}|^{2}+|z_{3}|^{2}=\tanh ^{2}\rho =1-|z_{1}|^{2}
\label{AldAdSZ-1-1-1-1-2-1}
\end{equation}%
suggests that there is an $S^{3}$ and its radius is not constant and changes
with $|z_{1}|$. This means that the size of the $S^{3}~$changes with a
radial direction, and the radius of $S^{3}$ and the $|z_{1}|$, i.e. the
radius of $S^{1}$, combine into the overall radial coordinate of the 6d
space. In other words, the radius of $S^{3}$ may be viewed as the radial
coordinate of the 6d space, projected to the $(z_{2},z_{3}$) subspace.

\subsection{1/2 BPS geometries}

\vspace{1pt}We can also embed the 1/2 BPS geometries as the excitations, in
the similar way as in e.g. subsection \ref{sec:1-8_SSJ_AdS_1st}. The $y$ in
this subsection denotes the $y_{1/2}~$of the 1/2 BPS ansatz. We have
\begin{equation}
Z(z_{1},\bar{z}_{1},y)=-\frac{1}{2}+\frac{{y^{2}}}{\pi }\int_{D}\frac{{%
dx_{1}^{\prime }dx_{2}^{\prime }}}{{[|z_{1}-z_{1}^{\prime }|^{2}+y^{2}]^{2}}}%
,  \label{eq:llmgreens}
\end{equation}%
where the integral is over the areas of the $Z=1/2$ droplets\ in the $z_{1}$%
~subspace. From (\ref{eq:7rydef}) we have
\begin{equation}
\log (r^{2})=-\frac{1}{\pi }\int_{D}\frac{{dx_{1}^{\prime }dx_{2}^{\prime }}%
}{{|z_{1}-z_{1}^{\prime }|^{2}+y^{2}}}.  \label{eq:7rygreens}
\end{equation}

Integrating (\ref{eq:llmgreens}) as in (\ref{eq:7kpotint}) gives an
expression for the Kahler potential
\begin{equation}
K=-\frac{1}{2\pi }\int_{D}\left( \frac{{y^{2}}}{{|z_{1}-z_{1}^{\prime
}|^{2}+y^{2}}}-\log [|z_{1}-z_{1}^{\prime }|^{2}+y^{2}]\right)
dx_{1}^{\prime }dx_{2}^{\prime }+\frac{1}{2}-\frac{1}{2}\log |z_{1}|^{2}.
\label{eq:7kgreens}
\end{equation}%
and $K$ in this case only depends on ${z_{1},}\bar{z}_{1}$ and\ $r$.

In the large $y_{1/2}$ region ,the expression of $y_{1/2}$ approaches
\begin{equation}
y_{1/2}^{2}=(\frac{1}{1-|z_{2}|^{2}-|z_{3}|^{2}}%
-|z_{1}|^{2})(|z_{2}|^{2}+|z_{3}|^{2}),
\end{equation}%
which implies that large $y_{1/2}$ corresponds to $|z_{2}|^{2}+|z_{3}|^{2}%
\rightarrow 1.$~At large $y_{1/2}$, we also have
\begin{eqnarray}
&&\frac{1}{2}\log (|z_{1}|^{2}+y_{1/2}^{2})\simeq -\frac{1}{2}\mathrm{\log }%
(1-|z_{2}|^{2}-|z_{3}|^{2}), \\
&&-\frac{1}{2}\frac{y_{1/2}^{2}}{{|z_{1}|^{2}+}y_{1/2}^{2}}\simeq -\frac{1}{2%
}+\frac{1}{2}(1-|z_{2}|^{2}-|z_{3}|^{2})|z_{1}|^{2}.
\end{eqnarray}

\subsection{More general cases}

Both the above the two cases in subsections \ref{sec:1-8_SSJ_AdS_1st}, \ref%
{sec:1-8_SSJ_AdS_2nd} for $AdS$ have the similar form, i.e.

\begin{equation}
K=\frac{1}{2}|z_{1}|^{2}(1-|z_{2}|^{2}-|z_{3}|^{2})-\frac{1}{2}\log
((1-|z_{2}|^{2}-|z_{3}|^{2})|z_{1}|^{2}),
\end{equation}%
for the first case and
\begin{equation}
K=\frac{1}{2}|z_{1}|^{-2}(1-|z_{2}|^{2}-|z_{3}|^{2})-\frac{1}{2}\log
((1-|z_{2}|^{2}-|z_{3}|^{2})|z_{1}|^{-2}).
\end{equation}%
for another case.

For more general geometries, one change of variable is

\begin{equation}
K=\frac{1}{2}f(z_{1},\bar{z}_{1})(1-s(z_{i},\bar{z}_{i}))-\frac{1}{2}\log
(1-s(z_{i},\bar{z}_{i}))-\frac{1}{2}\log f(z_{1},\bar{z}_{1})
\label{K_ge_SSJ}
\end{equation}%
where $i=2,3.$ This is also similar to (\ref{K_JJJ_ge_01}) in subsection \ref%
{sec:1-8_JJJ_more_general}.

One may also introduce the ansatz for adding ripples on the $S^{3}$
direction, e.g.%
\begin{equation}
s(z_{i},\bar{z}_{i})=(|z_{2}|^{2}+|z_{3}|^{2})\mathrm{exp}(\tilde{f}%
\sum_{l_{2},l_{3}\in Z^{+}}\epsilon
_{l_{2},l_{3}}(z_{2}^{l_{2}}z_{3}^{l_{3}}+c.c.)).
\end{equation}%
in (\ref{K_ge_SSJ}).

\subsection{Eigenvalue approach}

\label{sec:1-8_SSJ_AdS_1st_eigen}

\vspace{1pt} The subsections \ref{sec:1-4_SJ_AdS_1st}, \ref%
{sec:1-8_SSJ_AdS_1st} suggest that the droplet space of a class of 1/4 BPS
states with $S_{1},J~$and of a class of 1/8 BPS states with $S_{1},S_{2},J$
may be considered as $D^{2}\times R^{2},~B^{4}\times R^{2}$ respectively. In
this subsection, we study systems of eigenvalues in the space $%
B^{2n-2}\times R^{2},~n=2,3,4$, where $B^{2n-2}$ denotes a ball with $2n-2$
dimensions. The $B^{2n-2}~$may be viewed as $R^{2n-2}~$with a sphere at
infinity removed. One can also take a limit focusing on the region near the
origin of $B^{2n-2}$ and obtain $R^{2n-2}$ in this limiting procedure.

We first study the case $D^{2}\times R^{2}.$\ The distance between two
points on the disk, in the unit that the radius is 1, is $\rho
(u_{1},u_{2}), $~and
\begin{equation}
\sinh ^{2}\rho (u_{1},u_{2})=\frac{\left\vert u_{1}-u_{2}\right\vert ^{2}}{%
(1-\left\vert u_{1}\right\vert ^{2})(1-\left\vert u_{2}\right\vert ^{2})}
\end{equation}%
where $u$ is a complex coordinate and $\left\vert u\right\vert ^{2}\leqslant
1.~$The potential energy between particles can be written in the form that
is the solution to the Poisson equation on the disk,
\begin{equation}
\nabla ^{2}V=-2\pi \delta (\rho (u_{1},u_{2}))-2,
\end{equation}%
and one can obtain a solution%
\begin{equation}
V=-{\frac{1}{2}}\log \sinh ^{2}\rho (u_{1},u_{2}).
\end{equation}

So we see that the force between two particles is given by the potential
energy%
\begin{equation}
V_{j,k}=-{\frac{1}{2}}\eta _{k}\eta _{j}\log \sinh ^{2}\rho (u_{k},u_{j})
\end{equation}%
where $\eta _{k}$ is the charge of the particle.

For the case of $D^{2}$,~with coordinates $u_{2},\bar{u}_{2}$, the energy
for the system is%
\begin{equation}
\mathcal{H}_{eff}=-\frac{1}{2}N\sum_{k}\log (u_{0}^{2}-\left\vert
u_{2,k}\right\vert ^{2})-\frac{1}{2}\sum_{j<k}\log (\left\vert
u_{2,k}-u_{2,j}\right\vert ^{2})  \label{H_eff_S}
\end{equation}%
where $k$ labels individual eigenvalues, and $u_{0}$ is a real number, which
is the radius of the disk. In the case of $B^{4},$ with two complex
coordinates $u_{2},u_{3}$, we have a rotational symmetry between $%
u_{2},u_{3} $, so the effective Hamiltonian would be a generalization of (%
\ref{H_eff_S}),
\begin{equation}
\mathcal{H}_{eff}=-\frac{1}{2}N\sum_{k}\log (u_{0}^{2}-\left\vert
u_{2,k}\right\vert ^{2}-\left\vert u_{3,k}\right\vert ^{2})-\frac{1}{2}%
\sum_{j<k}\log (\left\vert u_{2,k}-u_{2,j}\right\vert ^{2}+\left\vert
u_{3,k}-u_{3,j}\right\vert ^{2})
\end{equation}%
where $u_{0}~$denotes the radius of $B^{4}.$ The last term is due to the
similarity with \cite{Berenstein:2005aa}.

For the case of $B^{4}\times R^{2},$ due to the symmetry for the measure
term, we can have that
\begin{eqnarray}
&&\mathcal{H}_{eff}=\frac{1}{2}{\sum_{k}}\left\vert u_{1,k}\right\vert ^{2}-%
\frac{1}{2}N\sum_{k}\log (1-\frac{\left\vert u_{2,k}\right\vert ^{2}}{%
u_{0}^{2}}-\frac{\left\vert u_{3,k}\right\vert ^{2}}{u_{0}^{2}})  \notag \\
&&-\frac{1}{2}\sum_{j<k}\log (\left\vert u_{1,k}-u_{1,j}\right\vert
^{2}+\left\vert u_{2,k}-u_{2,j}\right\vert ^{2}+\left\vert
u_{3,k}-u_{3,j}\right\vert ^{2})  \label{H_eff_SSJ}
\end{eqnarray}%
up to an overall constant shift, and $u_{1},\bar{u}_{1}~$denote the $R^{2}~$%
direction.

From the point of view of the wavefunction norm
\begin{equation}
\left\langle \psi \mid \psi \right\rangle \sim e^{-2\mathcal{H}_{eff}}
\end{equation}%
the factor $(u_{0}^{2}-\left\vert u_{2,k}\right\vert ^{2}-\left\vert
u_{3,k}\right\vert ^{2})$ will appear in the wavefunction norm, and
guarantee that it vanishes at the boundary of $B^{4},~$i.e.$~\left\vert
u_{2,k}\right\vert ^{2}+\left\vert u_{3,k}\right\vert ^{2}=u_{0}^{2}.~$We
can also introduce excited state wavefunctions by the multiplication of the
ground state wavefunction by additional factors, and change the $\mathcal{H}%
_{eff}$. Also, one can obtain an integral expression for the effective
Hamiltonian using an eigenvalue density function, replacing the sums. We can
also have a limit from (\ref{H_eff_SSJ}) when setting $u_{3,k}=0,~$which is
the $D^{2}\times R^{2}~$case$.$

Equation (\ref{H_eff_SSJ}) may also be generalized to higher dimensions for $%
B^{2n-2}\times R^{2},$%
\begin{equation}
\mathcal{H}_{eff}=\frac{1}{2}{\sum_{k}}\left\vert u_{1,k}\right\vert ^{2}-%
\frac{1}{2}N\sum_{k}\log (1-\frac{\sum_{i=2}^{n}\left\vert
u_{i,k}\right\vert ^{2}}{u_{0}^{2}})-\frac{1}{2}\sum_{j<k}\log
(\sum_{i=1}^{n}\left\vert u_{i,k}-u_{i,j}\right\vert ^{2})
\end{equation}%
where the last term is due to the similarity with \cite{Berenstein:2005aa}.
The discussion in this subsection might be applicable for higher $n,$ e.g. $%
B^{6}$ case which may be relevant for 6D theories.

We make a change of variable%
\begin{equation}
z_{1}=\frac{L}{\sqrt{N}}u_{1},\quad z_{2,3}=L\frac{u_{2,3}}{u_{0}},~~K=\frac{%
L^{2}}{N}H_{eff}(u_{i},\bar{u}_{i})
\end{equation}%
and (\ref{H_eff_SSJ}) becomes
\begin{eqnarray}
&&\frac{L^{2}}{N}\mathcal{H}_{eff}=\frac{1}{2}{\sum_{k}}\left\vert
z_{1,k}\right\vert ^{2}-\frac{1}{2}L^{2}\sum_{k}\log (1-\frac{\left\vert
z_{2,k}\right\vert ^{2}}{L^{2}}-\frac{\left\vert z_{3,k}\right\vert ^{2}}{%
L^{2}})  \notag \\
&&-\frac{1}{2}\frac{L^{2}}{N}\sum_{j<k}\log (\left\vert
z_{1,k}-z_{1,j}\right\vert ^{2}+\frac{u{}_{0}^{2}}{N}(\left\vert
z_{2,k}-z_{2,j}\right\vert ^{2}+\left\vert z_{3,k}-z_{3,j}\right\vert ^{2}))
\end{eqnarray}%
up to an overall constant shift $c=-\frac{L^{2}(N-1)}{4}\log \frac{N}{L^{2}}$
in this case.

Now we look at the test eigenvalue, and we have the effective Hamiltonian
for the test eigenvalue,
\begin{eqnarray}
&&\frac{L^{2}}{N}H_{eff}=\frac{1}{2}\left\vert z_{1}\right\vert ^{2}-\frac{1%
}{2}L^{2}\log (1-\frac{\left\vert z_{2}\right\vert ^{2}}{L^{2}}-\frac{%
\left\vert z_{3}\right\vert ^{2}}{L^{2}})  \notag \\
&&-\frac{1}{2}\frac{L^{2}}{N}\sum_{j}\log (\left\vert
z_{1}-z_{1,j}\right\vert ^{2}+\frac{u{}_{0}^{2}}{N}(\left\vert
z_{2}-z_{2,j}\right\vert ^{2}+\left\vert z_{3}-z_{3,j}\right\vert ^{2})) \\
&\simeq &\textstyle\frac{1}{2}\left\vert z_{1}\right\vert ^{2}-\frac{1}{2}%
L^{2}\log (1-\frac{\left\vert z_{2}\right\vert ^{2}}{L^{2}}-\frac{\left\vert
z_{3}\right\vert ^{2}}{L^{2}})-\frac{1}{2}L^{2}\log (\left\vert
z_{1}\right\vert ^{2}+\frac{u{}_{0}^{2}}{N}(\left\vert z_{2}\right\vert
^{2}+\left\vert z_{3}\right\vert ^{2})).  \label{H_eff_sum_over}
\end{eqnarray}%
In the last line, we assumed that the eigenvalue distribution has an $S^{1}~$%
symmetry in $z_{1}~$space and an $S^{3}~$symmetry in $z_{2},z_{3}~$space,
and used the approximation that the overall force for the test eigenvalue
would be exerted from the origin, in the leading order, similar to the the
approximation in (\ref{K_d_JJ}),(\ref{K_d_JJJ}).

Now we look at a limiting case
\begin{equation}
~\left\vert z_{2}\right\vert ^{2}+\left\vert z_{3}\right\vert ^{2}\ll
\left\vert z_{1}\right\vert ^{2}  \label{regime_01}
\end{equation}%
and $\left\vert z_{1}\right\vert ^{2}~$is finite. We also have $\frac{%
u{}_{0}^{2}}{N}$ finite. We then have from (\ref{H_eff_sum_over}),
\begin{eqnarray}
&&{\textstyle\frac{L^{2}}{N}}H_{eff}\simeq \textstyle\frac{1}{2}\left\vert
z_{1}\right\vert ^{2}(1-\frac{\left\vert z_{2}\right\vert ^{2}}{L^{2}}-\frac{%
\left\vert z_{3}\right\vert ^{2}}{L^{2}})-\textstyle\frac{1}{2}L^{2}\log (1-%
\frac{\left\vert z_{2}\right\vert ^{2}}{L^{2}}-\frac{\left\vert
z_{3}\right\vert ^{2}}{L^{2}})-\textstyle\frac{1}{2}L^{2}\log \left\vert
z_{1}\right\vert ^{2}+o(\left\vert z_{2}\right\vert ^{2}+\left\vert
z_{3}\right\vert ^{2}).  \notag \\
&&  \label{H_eff_ap}
\end{eqnarray}

The $K$ for $AdS$ in subsection \ref{sec:1-8_SSJ_AdS_1st} is
\begin{equation}
K=\textstyle{\frac{1}{2}\left\vert z_{1}\right\vert ^{2}(1-\frac{\left\vert
z_{2}\right\vert ^{2}}{L^{2}}-\frac{\left\vert z_{3}\right\vert ^{2}}{L^{2}}%
)-\frac{1}{2}L^{2}\log (1-\frac{\left\vert z_{2}\right\vert ^{2}}{L^{2}}-%
\frac{\left\vert z_{3}\right\vert ^{2}}{L^{2}})-\frac{1}{2}L^{2}\log
\left\vert z_{1}\right\vert ^{2}.}
\end{equation}%
We see that this and $\frac{L^{2}}{N}H_{eff}~$in (\ref{H_eff_ap})
approximately matches in the regime (\ref{regime_01}).

Now we consider (\ref{H_eff_SSJ}) in the region near the origin of $B^{4},~$%
i.e. when $\frac{\left\vert u_{2,k}\right\vert ^{2}}{u_{0}^{2}}{\small +}%
\frac{\left\vert u_{3,k}\right\vert ^{2}}{u_{0}^{2}}{\small \ll 1}$,
\begin{eqnarray}
&&\mathcal{H}_{eff}\simeq \frac{1}{2}{\sum_{k}}\left\vert u_{1,k}\right\vert
^{2}+\frac{1}{2}\sum_{k}\frac{N}{u_{0}^{2}}(\left\vert u_{2,k}\right\vert
^{2}+\left\vert u_{3,k}\right\vert ^{2})  \notag \\
&&-\frac{1}{2}\sum_{j<k}\log (\left\vert u_{1,k}-u_{1,j}\right\vert
^{2}+\left\vert u_{2,k}-u_{2,j}\right\vert ^{2}+\left\vert
u_{3,k}-u_{3,j}\right\vert ^{2}).
\end{eqnarray}%
In this case, we have $B^{4}\rightarrow R^{4}.~$We have the mass terms for $%
u_{2},u_{3},$ as given from $\frac{N}{u_{0}^{2}}.$ If we think of the $u_{1}$
as from the scalar $Z~$in $\mathcal{N}$=4 SYM, we argue that these other two
eigenvalues $u_{2},u_{3}$ may come from other fields, or a few higher modes
of $Z~$or other fields. The precise realization would depend on which sector
of the states that we are focusing on.

\subsection{Matrix model method}

In this subsection, we consider relating the states with $S_{1},S_{2},J$
described in subsection \ref{sec:1-8_SSJ_general_ansatz}, to the eigenvalues
from the $\mathcal{N}$=4 SYM.

We consider writing the scalar $Z$ as%
\begin{equation}
Z=\Phi _{1}+Y_{1,\frac{1}{2},\frac{1}{2}}(\Omega )\Phi _{2}+Y_{1,\frac{1}{2}%
,-\frac{1}{2}}(\Omega )\Phi _{3}
\end{equation}%
where $Y_{l,s_{L3},s_{R3}}(\Omega )$ denote orthonormalized scalar spherical
harmonics on $S^{3}$, in ($\frac{l}{2},\frac{l}{2}$) representation of SU(2)$%
_{L}\times $SU(2)$_{R},~$and $s_{L3},s_{R3}~$are the $j_{3}$ of each SU(2).
The $\Omega ~$denotes angular coordinates on the sphere, and the $\Phi
_{1},\Phi _{2},\Phi _{3}$ are complex matrices.

The action involving the sector we are considering is
\begin{equation}
S\simeq \int dt\frac{1}{2}\text{tr}\left( \left\vert D_{0}\Phi
_{1}\right\vert ^{2}-\left\vert \Phi _{1}\right\vert ^{2}+\left\vert
D_{0}\Phi _{2}\right\vert ^{2}+\left\vert D_{0}\Phi _{3}\right\vert
^{2}-4\left\vert \Phi _{2}\right\vert ^{2}-4\left\vert \Phi _{3}\right\vert
^{2}\right)
\end{equation}%
where we have integrated out angular coordinates on $S^{3}$, and then
rescaled the fields such that they have canonical kinetic terms as in the
last line.

We can consider the Hamiltonian%
\begin{equation}
\mathcal{H}=\frac{1}{2}\text{{tr}}\left( \left\vert D_{0}\Phi
_{1}\right\vert ^{2}+\left\vert \Phi _{1}\right\vert ^{2}+\left\vert
D_{0}\Phi _{2}\right\vert ^{2}+\left\vert D_{0}\Phi _{3}\right\vert
^{2}+4\left\vert \Phi _{2}\right\vert ^{2}+4\left\vert \Phi _{3}\right\vert
^{2}\right) .  \label{Hamiltonian_SSJ}
\end{equation}%
One can add more fields, corresponding to larger sectors in the Hilbert
space. One can also consider interaction terms or including non-BPS states.

We may map the operators of the 4d theory to the wavefunctions in the
reduced 1d model, e.g. {\large
\begin{equation}
\prod_{i=1}^{m}\text{tr}(D_{+\dot{+}}^{n_{2i}}{}D_{+\dot{-}%
}^{n_{3i}}{}Z^{n_{1i}})\rightarrow \prod_{i=1}^{m}\text{tr}(\Phi
_{2}{}^{n_{2i}}\Phi _{3}{}^{n_{3i}}\Phi _{1}{}^{n_{1i}}).
\end{equation}%
}

One can also reduce the model (\ref{Hamiltonian_SSJ}) in the eigenvalue
basis, and then study the quantum mechanics of $N$ eigenvalues in 6d. Their
wavefunction norm defines the effective Hamiltonian,%
\begin{equation}
\left\langle \psi \mid \psi \right\rangle \sim e^{-2\mathcal{H}_{eff}}.
\end{equation}

The ground state wavefunction gives an effective Hamiltonian%
\begin{equation}
\mathcal{H}_{eff}=\frac{1}{2}{\sum_{k}}\left( \left\vert u_{1,k}\right\vert
^{2}+4(\left\vert u_{2,k}\right\vert ^{2}+\left\vert u_{3,k}\right\vert
^{2})\right) -\frac{1}{2}\sum_{j<k}\log (\left\vert
u_{1,k}-u_{1,j}\right\vert ^{2}+\left\vert u_{2,k}-u_{2,j}\right\vert
^{2}+\left\vert u_{3,k}-u_{3,j}\right\vert ^{2})
\label{H_eff_eigenvalue_SSJ}
\end{equation}%
where the complex numbers $u_{i,k}~$($k$ labels the $N$ eigenvalues) denote
the eigenvalues of the three matrices $\Phi _{i}$, and the last term is due
to the insertion of a measure factor in the wavefunction, and was analyzed
in \cite{Berenstein:2005aa} for multiple matrices at strong coupling. One
can also consider other excited state wavefunctions, which give other
effective Hamiltonians.

The effective Hamiltonian also gives a most probable distribution of
eigenvalues, when the eigenvalues approximately reaches the equilibrium
configuration. For example, from (\ref{H_eff_eigenvalue_SSJ}), the
eigenvalue density distribution would have an $S^{3}\times S^{1}$ symmetry.

In the continuum limit of the eigenvalue distribution, one can also
approximate the sums in (\ref{H_eff_eigenvalue_SSJ}) by integrals weighted
by eigenvalue density function $\rho (u_{i},\bar{u}_{i}),$ so
\begin{eqnarray}
&&\mathcal{H}_{eff}=\frac{1}{2}{\int d^{6}u}\rho (u_{i},\bar{u}%
_{i})(\left\vert u_{1}\right\vert ^{2}+4(\left\vert u_{2}\right\vert
^{2}+\left\vert u_{3}\right\vert ^{2}))  \notag \\
&&-\frac{1}{4}{\int \int d^{6}u}{d^{6}u^{\prime }}\rho (u_{i},\bar{u}%
_{i})\rho (u_{i}^{\prime },\bar{u}_{i}^{\prime })\log (\left\vert
u_{1}-u_{1}^{\prime }\right\vert ^{2}+\left\vert u_{2}-u_{2}^{\prime
}\right\vert ^{2}+\left\vert u_{3}-u_{3}^{\prime }\right\vert ^{2})+\sigma ({%
\int d^{6}u}\rho (u_{i},\bar{u}_{i})-N)  \notag \\
&&
\end{eqnarray}%
where we use the notation $d^{6}u=d^{2}u_{1}d^{2}u_{2}d^{2}u_{3},$ and $%
d^{2}u_{i}=\frac{1}{2i}du_{i}d\bar{u}_{i}=dx_{1}dy_{1}$. The last term is to
enforce the total number of eigenvalues when $\delta _{\sigma }\mathcal{H}%
_{eff}=0$.

\vspace{1pt}We can also reduce (\ref{H_eff_eigenvalue_SSJ}) or (\ref%
{Hamiltonian_SSJ}) to the case with the first two matrices $\Phi _{1},\Phi
_{2}~$only, and it would be relevant to the states with $S_{1},J$ as studied
in section \ref{sec:1-4_SJ}.

We see that the effective Hamiltonian of a test eigenvalue can be considered
as%
\begin{equation}
H_{test}=\delta _{\rho }\mathcal{H}_{eff}-\sigma ,
\end{equation}%
where $\delta _{\rho }\mathcal{H}_{eff}$ is the derivative of the effective
Hamiltonian $\mathcal{H}_{eff}~$with respect to the eigenvalue density. So
another interpretation of the $K_{d}$ as in subsections \ref%
{sec:1-4_JJ_non_radial}, \ref{sec:1-8_JJJ_AdS_eigen}, \ref%
{sec:1-8_SSJ_AdS_1st_eigen} would be $\delta _{\rho }\mathcal{H}%
_{eff}-\sigma $. We can also set $\sigma =0~$after setting the variation $%
\delta _{\sigma }\mathcal{H}_{eff}=0.$

Now we discuss the eigenvalue distribution. We make variation $\delta _{\rho
}\mathcal{H}_{eff},$%
\begin{equation}
\delta _{\rho }\mathcal{H}_{eff}=\frac{1}{2}(\left\vert u_{1}\right\vert
^{2}+4(\left\vert u_{2}\right\vert ^{2}+\left\vert u_{3}\right\vert ^{2}))-%
\frac{1}{2}{\int }{d^{6}u^{\prime }}\rho (u_{i}^{\prime },\bar{u}%
_{i}^{\prime })\log (\left\vert u_{1}-u_{1}^{\prime }\right\vert
^{2}+\left\vert u_{2}-u_{2}^{\prime }\right\vert ^{2}+\left\vert
u_{3}-u_{3}^{\prime }\right\vert ^{2})+\sigma =0.  \label{dH_eff_SSJ}
\end{equation}

The distribution would have an $S^{3}\times S^{1}$ symmetry. So we may assume%
\begin{equation}
\rho (u_{i},\bar{u}_{i})=\rho (\tilde{r},\left\vert u_{1}\right\vert )
\end{equation}%
where $\tilde{r}^{2}=\left\vert u_{2}\right\vert ^{2}+\left\vert
u_{3}\right\vert ^{2}.~$

We can make variation of (\ref{dH_eff_SSJ}) with respect to $u_{i},\bar{u}%
_{i},$ and obtain the equations for equilibrium configurations. For the
equilibrium configuration of the density distribution at low energies, we
have an $S^{3}$ symmetry in the $u_{2},u_{3}~$space. If the mass terms for $%
u_{2},u_{3}~$and for $u_{1}$ were equal, we would have a round $S^{5}$\cite%
{Berenstein:2005aa}.~One can consider this equilibrium configuration (\ref%
{dH_eff_SSJ}) as deforming the round $S^{5}~$due to increasing the mass
terms for two complex directions $u_{2},u_{3}$. The larger mass terms for $%
u_{2},u_{3}~$make the distribution on $S^{5}$ no longer uniform in all
angles, and they squeeze the $S^{3}$ directions.$~$We argue that the
eigenvalue distribution form deformed $S^{5}~$with $S^{3}\times S^{1}~$%
symmetry, with the $S^{3}$ directions squeezed or decreased in size.

The $S^{3}$ here is the $S^{3}$ in $AdS$ directions, and for the ground
state configuration, we have a round $S^{3}~$symmetry in the eigenvalue
space.~For other 1/8 BPS geometries, one can add ripples on this $S^{3}.~$We
have therefore argued a possibility that the $S^{3}$ symmetry may be related
to the distribution of the eigenvalues of the $\Phi _{2},\Phi _{3}.$

\section{Discussion}

\label{sec:discussion}

In this paper, we studied four sectors of geometries corresponding to
various sectors of 1/4 BPS and 1/8 BPS states in $\mathcal{N}$=4 SYM, with
angular momenta in $S^{5}~$directions as well as spins in $AdS_{5}~$%
directions.

For the states with angular momenta $J_{1},J_{2}~$in $S^{5}~$directions, we
first analyzed the small $y$ equations, and then studied the coupled
equations for $K_{0},K_{1}~$in the $Z$=1/2 region in some detail. We also
studied changes of variables, e.g. (\ref{ansatz_non_radial}), (\ref%
{change_variable_JJ}), which change the Monge-Ampere type equation into
other equations, and we studied them in some detail. We also studied the
features of adding ripples to the droplets in the $y$=0 hyperplane.

For the states with $J_{1},J_{2},J_{3}$ in $S^{5}~$directions, we studied
similar issues, including, among other things, analyzing their relations to
the eigenvalue pictures.

For the states with spin $S_{1}~$in $AdS_{5}~$and $J~$in $S^{5}~$directions
respectively, we first studied multiple embeddings of $AdS$, and then
analyzed small $y$ and large $y$ behaviors of more general solutions. Also,
we discussed the inversion symmetry occurred e.g. in the embeddings of $AdS$.

For the states with spins $S_{1},S_{2}~$in $AdS_{5}~$and $J~$in $S^{5}~$%
directions respectively, we studied similar issues, including, among other
things, the multiple embeddings and inversion. We also studied the gauge
theory side briefly.

It would be nice to understand many aspects in these topics better in the
general context of the AdS/CFT correspondence \cite{Aharony:1999ti}.

\vspace{1pt}

\section*{Acknowledgments}

This work is supported in part by the ME and Feder (grant FPA2008- 01838),
by the Spanish Consolider-Ingenio 2010 Programme CPAN (CSD2007-00042), by
the Juan de la Cierva program of MCyI of Spain, and by Xunta de Galicia
(Conselleria de Educacion and grants PGIDIT06 PXIB206185Pz and INCITE09 206
121 PR). We also would like to thank J. P. Shock and J. Tarrio for
correspondences.


\appendix



\begin{thebibliography}{99}
\bibitem{Berenstein:2005aa} D.~Berenstein,
JHEP 0601, 125 (2006) [arXiv:hep-th/0507203].


\bibitem{Corley:2001zk} S.~Corley, A.~Jevicki and S.~Ramgoolam,
Adv.\ Theor.\ Math.\ Phys.\ 5, 809 (2002) [arXiv:hep-th/0111222].


\bibitem{Berenstein:2004kk} D.~Berenstein,
JHEP 0407, 018 (2004) [arXiv:hep-th/0403110].


\bibitem{Lin:2004nb} H.~Lin, O.~Lunin and J.~M.~Maldacena,
JHEP 0410, 025 (2004) [arXiv:hep-th/0409174].


\bibitem{Chen:2007gh} H.~Y.~Chen, D.~H.~Correa and G.~A.~Silva,
Phys.\ Rev.\ D 76, 026003 (2007) [arXiv:hep-th/0703068].


\bibitem{Vazquez:2006id} S.~E.~Vazquez, {}
Phys.\ Rev.\ D 75, 125012 (2007) [arXiv:hep-th/0612014].


\bibitem{Correa:2010zj} D.~H.~Correa and M.~Wolf,
arXiv:1007.5284 [hep-th]. 

\bibitem{Berenstein:2007wz} D.~Berenstein and R.~Cotta,
JHEP 0704, 071 (2007) [arXiv:hep-th/0702090].


\bibitem{Koch:2008ah} R.~d.~M.~Koch, 
JHEP 0811, 061 (2008) [arXiv:0806.0685 [hep-th]].


\bibitem{Lin:2010sb} H.~Lin, A.~Morisse and J.~P.~Shock,
JHEP 1006, 055 (2010) [arXiv:1003.4190 [hep-th]].


\bibitem{Gukov:2008sn} S.~Gukov and E.~Witten,
arXiv:0804.1561 [hep-th]. 


\bibitem{Gomis:2007fi} J.~Gomis and S.~Matsuura,
JHEP 0706, 025 (2007) [arXiv:0704.1657 [hep-th]].

\bibitem{Lin:2005nh} H.~Lin and J.~M.~Maldacena,
Phys.\ Rev.\ D 74, 084014 (2006) [arXiv:hep-th/0509235].


\bibitem{Grant:2005qc} L.~Grant, L.~Maoz, J.~Marsano, K.~Papadodimas and
V.~S.~Rychkov,
JHEP 0508, 025 (2005) [arXiv:hep-th/0505079].


\bibitem{Mandal:2005wv} G.~Mandal,
JHEP 0508, 052 (2005) [arXiv:hep-th/0502104].


\bibitem{Takayama:2005yq} Y.~Takayama and A.~Tsuchiya,
JHEP 0510, 004 (2005) [arXiv:hep-th/0507070].


\bibitem{Yoneya:2005si} T.~Yoneya,
JHEP \textbf{0512}, 028 (2005) [arXiv:hep-th/0510114].


\bibitem{Caldarelli:2004mz} M.~M.~Caldarelli, D.~Klemm and P.~J.~Silva,
Class.\ Quant.\ Grav.\ 22, 3461 (2005) [arXiv:hep-th/0411203].


\bibitem{Brown:2006zk} T.~W.~Brown, R.~de Mello Koch, S.~Ramgoolam and
N.~Toumbas, 
JHEP 0703, 072 (2007) [arXiv:hep-th/0611290].


\bibitem{Horava:2005pv} P.~Horava and P.~G.~Shepard,
JHEP 0502, 063 (2005) [arXiv:hep-th/0502127].


\bibitem{Mukhi:2005cv} S.~Mukhi and M.~Smedback,
JHEP 0508, 005 (2005) [arXiv:hep-th/0506059].


\bibitem{Shieh:2007xn} H.~H.~Shieh, G.~van Anders and M.~Van Raamsdonk,
JHEP 0709, 059 (2007) [arXiv:0705.4308 [hep-th]].

\bibitem{Horowitz:2006ct} G.~T.~Horowitz and J.~Polchinski,
arXiv:gr-qc/0602037. 


\bibitem{Balasubramanian:2005mg} V.~Balasubramanian, J.~de Boer, V.~Jejjala
and J.~Simon,
JHEP 0512, 006 (2005) [arXiv:hep-th/0508023].


\bibitem{Suryanarayana:2004ig} N.~V.~Suryanarayana,
JHEP 0601, 082 (2006) [arXiv:hep-th/0411145].

\bibitem{Balasubramanian:2007zt} V.~Balasubramanian, B.~Czech, K.~Larjo,
D.~Marolf and J.~Simon,
JHEP 0712, 067 (2007) [arXiv:0705.4431 [hep-th]].


\bibitem{Chong:2004ce} Z.~W.~Chong, H.~Lu and C.~N.~Pope,
Phys.\ Lett.\ B 614, 96 (2005) [arXiv:hep-th/0412221].


\bibitem{Donos:2006iy} A.~Donos,
Phys.\ Rev.\ D 75, 025010 (2007) [arXiv:hep-th/0606199].


\bibitem{Kim:2005ez} N.~Kim,
JHEP 0601, 094 (2006) [arXiv:hep-th/0511029].


\bibitem{Chen:2007du} B.~Chen, S. Cremonini, A. Donos, F.-L. Lin, H. Lin, J.
T. Liu, D. Vaman and W.-Y. Wen,
JHEP 0710, 003 (2007) [arXiv:0704.2233 [hep-th]].


\bibitem{Gava:2006pu} E.~Gava, G.~Milanesi, K.~S.~Narain and M.~O'Loughlin,
JHEP 0705, 030 (2007) [arXiv:hep-th/0611065].


\bibitem{Donos:2006ms} A.~Donos,
JHEP 0705, 072 (2007) [arXiv:hep-th/0610259].


\bibitem{Lunin:2008tf} O.~Lunin, 
JHEP 0809, 028 (2008) [arXiv:0802.0735 [hep-th]].

\bibitem{Gauntlett:2007ts} J.~P.~Gauntlett and N.~Kim,
Commun.\ Math.\ Phys.\ 284, 897 (2008) [arXiv:0710.2590 [hep-th]].

\bibitem{Liu:2004ru} J.~T.~Liu, D.~Vaman and W.~Y.~Wen,
Nucl.\ Phys.\ B 739, 285 (2006) [arXiv:hep-th/0412043].

\bibitem{Gauntlett:2006ns} J.~P.~Gauntlett, N.~Kim and D.~Waldram,
JHEP \textbf{0704}, 005 (2007) [arXiv:hep-th/0612253].


\bibitem{Biswas:2006tj} I.~Biswas, D.~Gaiotto, S.~Lahiri and S.~Minwalla,
JHEP 0712, 006 (2007) [arXiv:hep-th/0606087].


\bibitem{Mandal:2006tk} G.~Mandal and N.~V.~Suryanarayana,
JHEP 0703, 031 (2007) [arXiv:hep-th/0606088].


\bibitem{Yamaguchi:2006te} S.~Yamaguchi,
Int.\ J.\ Mod.\ Phys.\ A 22, 1353 (2007) [arXiv:hep-th/0601089].


\bibitem{Lunin:2006xr} O.~Lunin,
JHEP 0606, 026 (2006) [arXiv:hep-th/0604133].


\bibitem{Yamaguchi:2006tq} S.~Yamaguchi,
JHEP 0605, 037 (2006) [arXiv:hep-th/0603208].


\bibitem{Gomis:2006sb} J.~Gomis and F.~Passerini,
JHEP 0608, 074 (2006) [arXiv:hep-th/0604007].

\bibitem{Gomis:2006cu} J.~Gomis and C.~Romelsberger,
JHEP 0608, 050 (2006) [arXiv:hep-th/0604155].

\bibitem{Drukker:2006zk} N.~Drukker, S.~Giombi, R.~Ricci and D.~Trancanelli,
JHEP 0704, 008 (2007) [arXiv:hep-th/0612168].


\bibitem{Okuyama:2006jc} K.~Okuyama and G.~W.~Semenoff,
JHEP 0606, 057 (2006) [arXiv:hep-th/0604209].


\bibitem{Hartnoll:2006is} S.~A.~Hartnoll and S.~P.~Kumar,
JHEP 0608, 026 (2006) [arXiv:hep-th/0605027].


\bibitem{D'Hoker:2007fq} E.~D'Hoker, J.~Estes and M.~Gutperle,
JHEP 0706, 063 (2007) [arXiv:0705.1004 [hep-th]].

%

\bibitem{Aharony:1999ti} O.~Aharony, S.~S.~Gubser, J.~M.~Maldacena,
H.~Ooguri and Y.~Oz,
Phys.\ Rept.\ 323, 183 (2000) [arXiv:hep-th/9905111].


\bibitem{Brown:2007xh} T.~W.~Brown, P.~J.~Heslop and S.~Ramgoolam,
JHEP 0802, 030 (2008) [arXiv:0711.0176 [hep-th]].


\bibitem{Kimura:2007wy} Y.~Kimura and S.~Ramgoolam,
JHEP 0711, 078 (2007) [arXiv:0709.2158 [hep-th]].


\bibitem{Bhattacharyya:2008rb} R.~Bhattacharyya, S.~Collins and
R.~d.~M.~Koch, 
JHEP 0803, 044 (2008) [arXiv:0801.2061 [hep-th]].


\bibitem{Koch:2008cm} R.~d.~M.~Koch, N.~Ives and M.~Stephanou,
Phys.\ Rev.\ D 79, 026004 (2009) [arXiv:0810.4041 [hep-th]].


\bibitem{deMelloKoch:2009jc} R.~de Mello Koch, T.~K.~Dey, N.~Ives and
M.~Stephanou, 
JHEP 0908, 083 (2009) [arXiv:0905.2273 [hep-th]].
\end{thebibliography}
\end{document}